\documentclass[aps,prl,10pt,amsmath,amssymb,twocolumn,footinbib]{revtex4-1}

\usepackage[breaklinks,colorlinks=true,urlcolor=blue,citecolor=blue,linkcolor=blue]{hyperref}
\usepackage{color, braket}
\usepackage{graphicx}
\usepackage[large,bf]{subfigure}     

\newcommand{\comment}[1]{{}}
\definecolor{orange}{rgb}{1,0.5,0}

\newcommand{\hc}{\mathrm{H.c.}}

\usepackage{bbm}

\begin{document}

\title{Entanglement entropy of composite Fermi liquid states on the lattice:\\In support of the Widom formula}
\date{\today}
\author{Ryan V. Mishmash}
\author{Olexei I. Motrunich}
\affiliation{Department of Physics and Institute for Quantum Information and Matter, California Institute of Technology, Pasadena, CA 91125, USA}
\affiliation{Walter Burke Institute for Theoretical Physics, California Institute of Technology, Pasadena, CA 91125, USA}


\begin{abstract}
Quantum phases characterized by surfaces of gapless excitations are known to violate the otherwise ubiquitous boundary law of entanglement entropy in the form of a multiplicative log correction:  $S\sim L^{d-1} \log L$.  Using variational Monte Carlo, we calculate the second R\'enyi entropy for a model wavefunction of the $\nu=1/2$ composite Fermi liquid (CFL) state defined on the two-dimensional triangular lattice.  By carefully studying the scaling of the total R\'enyi entropy and, crucially, its contributions from the modulus and sign of the wavefunction on various finite-size geometries, we argue that the prefactor of the leading $L \log L$ term is equivalent to that in the analogous free fermion wavefunction.  In contrast to the recent results of Shao et al.~[PRL {\bf 114}, 206402 (2015)], we thus conclude that the ``Widom formula'' holds even in this non-Fermi liquid CFL state.  More generally, our results further elucidate---and place on a more quantitative footing---the relationship between nontrivial wavefunction sign structure and $S\sim L \log L$ entanglement scaling in such highly entangled gapless phases.
\end{abstract}

\maketitle

\begin{figure}[t]
\centerline{\includegraphics[width=0.97\columnwidth]{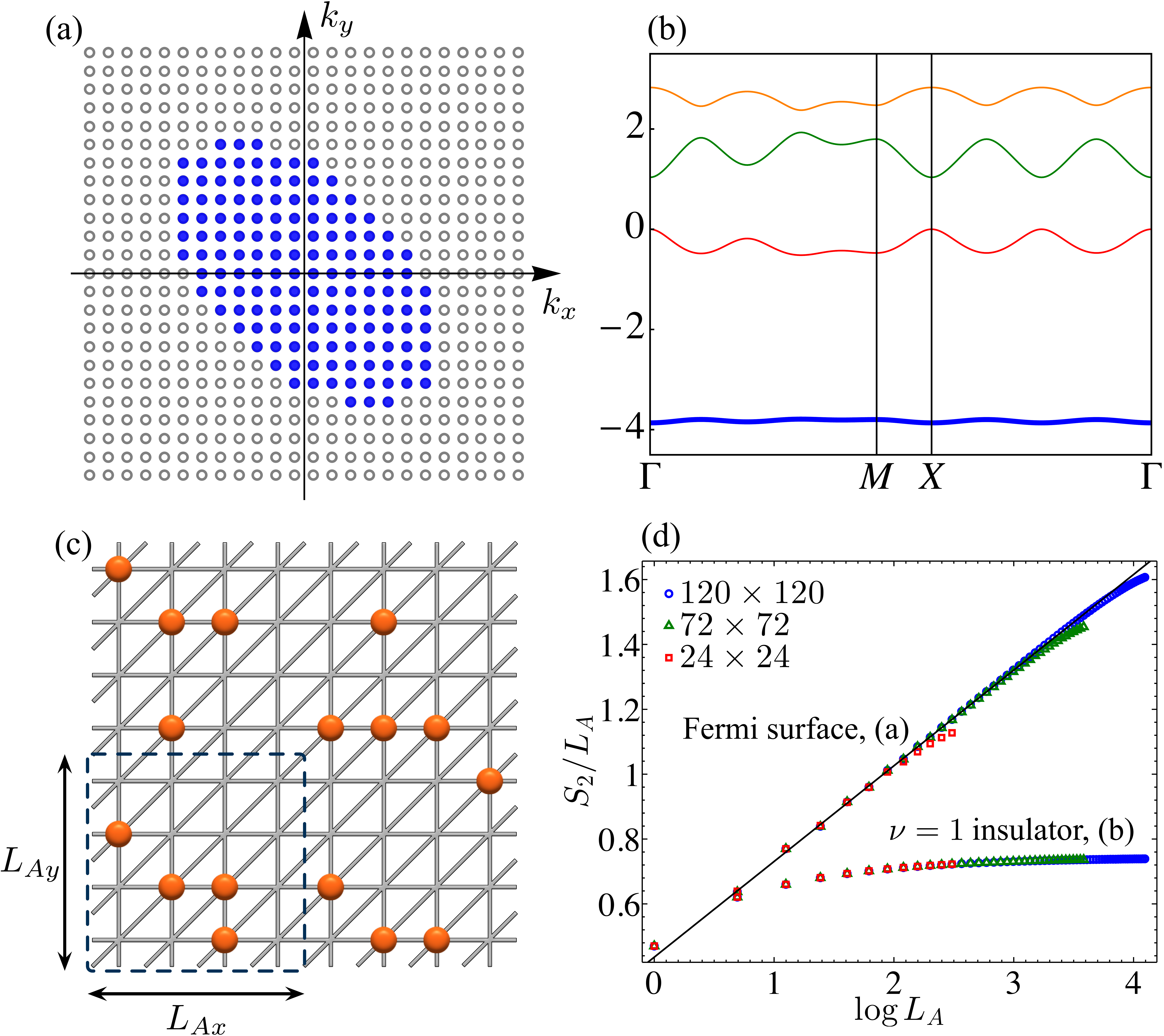}}
\caption{(a) Filled Fermi sea with a sharp Fermi surface used to construct $\Psi_f^\mathrm{FS}$ on a $24\times24$ lattice with $N=144$ electrons.  (b) Band structure for $d_{1,2}$; the $\Psi_{d_{1,2}}^{(\nu=1)}$ Slater determinants are constructed by filling the lowest, nearly flat band (blue) which has Chern number $C=1$.  (c) We work on the 2D triangular lattice 
and consider subregions of size $L_{Ax}\times L_{Ay}$ for our calculations of $S_2$ \cite{Note2}.  (d) $S_2$ scaling for the free fermion states in (a) and (b) for $L_{A}\times L_{A}$ subregions embedded in various $L\times L$ systems at $\rho=1/4$ (see legend); the black line indicates the Widom formula slope $\kappa_W$ (see text).}
\label{fig:Fig1}
\end{figure}

In recent years, bipartite entanglement entropy has emerged as an indispensable tool in the study of quantum many-body states \cite{Horodecki09_RMP_81_865, Laflorencie15_EEreview_arXiv_1512.03388}.  It can reveal highly universal, even nonlocal, information about a quantum phase given a ground state wavefunction.  While entanglement entropy has had remarkable success for gapped phases exhibiting topological order \cite{Kitaev06_TEE_PRL_96_110404, Levin06_TEE_PRL_96_110405, Zhang11_PRB_84_075128, Zhang12_MES_PRB_85_235151} and gapless Luttinger liquids \cite{Cardy04_JStatMech_P06002}, an interesting question concerns its ability to characterize two-dimensional (2D) \emph{highly entangled} systems containing a surface of gapless excitations in momentum space.  These states are known to exhibit a multiplicative log violation of the boundary law \cite{Wolf06_PRL_96_010404}:
\begin{equation}
S =  \kappa\,L_A\log L_A\, ,
\label{eq:LlogL}
\end{equation}
where $S$ is the entanglement entropy between a large real-space subregion of characteristic length $L_A$ and its complement (see Fig.~\ref{fig:Fig1}).

The free Fermi gas with a sharp Fermi surface is the simplest example of such a system [see Fig.~\ref{fig:Fig1}(a)].  In real space, however, the free fermion wavefunction is highly nontrivial, exhibiting complicated \emph{sign structure} \cite{Ceperley91_JStatPhys_63_1237} which is believed to be closely related to the anomalously large entanglement present in Eq.~\eqref{eq:LlogL}.  For free fermions, the coefficient $\kappa$ depends only on the \emph{shapes} of the subregion and Fermi surface and is given by an elegant geometric integral expression commonly referred to as the ``Widom formula'' \cite{Gioev06_PRL_96_100503, Barthel06_PRA_74_022329, Swingle10_PRL_105_050502, Spitzer14_WidomProof_PRL_112_160403}.

In fact, $\kappa$ is expected to be surprisingly universal and given by the Widom result $\kappa_W$ even for an interacting Fermi liquid \cite{Seidel12_PRX_2_011012, Swingle12_PRB_86_035116, Swingle12_RenyiWidom_PRB_86_045109}, as well as for more exotic states with emergent surfaces of gapless excitations \cite{Grover11_PRL_107_067202, Swingle13_PRB_87_045123, Swingle13_PRL_111_100405} which have the same Fermi surface content as the corresponding free Fermi gas.
Loosely speaking, $\kappa$ can thus generally be interpreted as measuring the ``gaplessness'' of the quantum state as contributed by the critical surface(s), emergent or otherwise \cite{Swingle10_PRL_105_050502, Swingle12_PRB_86_035116, Swingle12_RenyiWidom_PRB_86_045109, Swingle13_PRB_87_045123, Swingle13_PRL_111_100405, Lai13_BoseMetalEE_PRL_111_210402, Lai16_BoseVsFermi_PRB_93_121109}.

At present, several interesting open questions remain that we set out to address in this paper.  Which types of wavefunctions may violate the Widom formula?  More precisely, can a (possibly nonperturbatively strongly interacting) wavefunction with identical critical surfaces as the free Fermi gas have an entanglement scaling with $\kappa\neq\kappa_W$?  Since $\kappa=\kappa_W$ is expected to hold for interacting Fermi liquids \cite{Seidel12_PRX_2_011012, Swingle12_PRB_86_035116, Swingle12_RenyiWidom_PRB_86_045109}, can measuring $\kappa$ in a numerical simulation thus serve as a long-sought-after positive indicator of non-Fermi liquid behavior \cite{Kim15_PRL_114_206402}?  Finally, in practice, what is the best way to detect Widom-formula violation in numerical studies given the well-known signal-to-noise ratio problems inherent in Monte Carlo measurements of the entanglement entropy on large systems?

We now turn to the composite Fermi liquid (CFL) phase of the half-filled Landau level ($\nu=1/2$).  The CFL still stands today as the paradigmatic example of a strongly interacting gapless non-Fermi liquid state \cite{Halperin93_PRB_47_7312, Polchinski94_NuclPhysB_422_617, Nayak94_NuclPhysB_417_359, Nayak94_NuclPhysB_430_534, Altshuler94_PRB_50_14048, Kim95_PRB_52_17275} (see also Refs.~\cite{Son15_DiracCFL_PRX_5_031027, Senthil15_DualDirac_PRX_5_041031, Metlitski15_ParticleVortex_arXiv_1505.05142, Senthil16_Dualities_PRB_93_085110, Geraedts16_CFLDMRG_Science_352_197, Murthy16_nu12Hami_PRB_93_085405, Mulligan15_MirrorCFL_PRB_92_235105, Mulligan15_phCFL_PRB_92_165125, Mross15_QEDduality_arXiv_1510.08455, Mulligan16_EmergentphCFL_arXiv_1603.05656, Jain16_newCFL_arXiv_1604.03911, Senthil16_newCFL_arXiv_1604.06807} for several recent exciting developments).  Following Halperin, Lee, and Read (HLR) \cite{Halperin93_PRB_47_7312}, a model wavefunction for the CFL reads \cite{Rezayi94_nu1half_PRL_72_900, Rezayi00_nu12_PRL_84_4685, Barkeshli12_PRB_86_075136}
\begin{equation}
\Psi_\mathrm{HLR}\left(\{\mathbf{r}_i\}\right) = \Psi_b^{(\nu=1/2)}\left(\{\mathbf{r}_i\}\right) \Psi_f^\mathrm{FS}\left(\{\mathbf{r}_i\}\right),
\label{eq:HLRwf}
\end{equation}
where $\Psi_b^{(\nu=1/2)}$ is a Laughlin-type wavefunction for bosons at $\nu=1/2$ \cite{Laughlin83_PRL_50_1395, Laughlin87_PRL_59_2095}, $\Psi_f^\mathrm{FS}$ is a wavefunction for fermions in zero field exhibiting a Fermi surface (FS), and $\{\mathbf{r}_i\}$ are the coordinates of the $N$ electrons at which both $\Psi_b^{(\nu=1/2)}$ and $\Psi_f^\mathrm{FS}$ are to be evaluated.

Recently, Ref.~\cite{Kim15_PRL_114_206402} presented a numerical study of the second R\'enyi entropy $S_2$ \footnote{For the purpose of studying the scaling in Eq.~\eqref{eq:LlogL}, $S_2$ should be equally effective as, say, the von Neumann entropy $S_1$ \cite{Swingle12_RenyiWidom_PRB_86_045109}.} for a continuum wavefunction in the form of Eq.~\eqref{eq:HLRwf} projected into the lowest Landau level on the torus.  These authors found that for square $L_A\times L_A$ subregions the prefactor $\kappa$ in the leading $L_A\log L_A$ term of $S_2$ is approximately \emph{twice} the corresponding Widom formula result, i.e., twice what is obtained for the zero-field free fermion wavefunction $\Psi_f^\mathrm{FS}$.  This is a very striking result.  Since $\Psi_b^{(\nu=1/2)}$ is a fully gapped state with a clear boundary law \cite{Zhang11_PRB_84_075128} (albeit a wavefunction with interesting structure of zeros and complex phases) and the Guztwiller projection implicit in Eq.~\eqref{eq:HLRwf} generally only tends to (slightly) decrease entanglement \cite{Grover11_PRL_107_067202, Sheng09_PRB_79_205112, Jiang13_Nature_493_39}, such a dramatic increase in $\kappa$ for this wavefunction is very unexpected and, if correct, could point to new physics at play which is currently not understood. 

Here, we study the entanglement entropy of analogous HLR-type wavefunctions on the lattice, which to our knowledge have not been considered before in detail in any capacity.  Our wavefunctions are particularly easy to define and straightforward to handle using variational Monte Carlo \cite{Ceperley77_PRB_16_3081, Gros89_AnnPhys_189_53, Grover11_PRL_107_067202}, yet they should be in the same quantum phase as the state considered in Ref.~\cite{Kim15_PRL_114_206402}.
We consider $N$ spinless electrons moving on a toroidal 2D triangular lattice [see Fig.~\ref{fig:Fig1}(c)] of dimension $L_x\times L_y$ with uniform magnetic flux penetrating the sample \cite{Hofstadter76_PRB_14_2239}.  For concreteness, we take an electron density $\rho=N/(L_x L_y)=1/4$ with $\pi/2$ external magnetic flux per triangle.  Our model HLR wavefunction for this $\nu=1/2$ system reads
\begin{equation}
\Psi_\mathrm{HLR}^\mathrm{ferm}\left(\{\mathbf{r}_i\}\right) = \Psi_{d_1}^{(\nu=1)}\left(\{\mathbf{r}_i\}\right) \Psi_{d_2}^{(\nu=1)}\left(\{\mathbf{r}_i\}\right) \Psi_f^\mathrm{FS}\left(\{\mathbf{r}_i\}\right).
\label{eq:fermHLR}
\end{equation}
(See Fig.~\ref{fig:Fig1} and \footnote{See the Supplemental Material for details of our projected wavefunctions and Monte Carlo simulations.} for details.)
Within a ``parton'' approach \cite{Wen99_PartonQH_PRB_60_8827, Lee06_RevModPhys_78_17}, Eq.~\eqref{eq:fermHLR} corresponds to decomposing the physical electron as $c=d_1 d_2 f$ subject to the constraint $d_1^\dagger d_1 = d_2^\dagger d_2 = f^\dagger f = c^\dagger c$ at each site.  We will also consider a bosonic analog of the HLR state appropriate for bosons at $\nu=1$ \cite{Haldane98_nu1bosons_NuclPhysB_516_719, Read98_nu1bosons_PRB_58_16262}.  The construction parallels the fermionic state of Eq.~\eqref{eq:fermHLR} with a final wavefunction given by $\Psi_\mathrm{HLR}^\mathrm{bos}\left(\{\mathbf{r}_i\}\right) = \Psi_{d_1}^{(\nu=1)}\left(\{\mathbf{r}_i\}\right) \Psi_f^\mathrm{FS}\left(\{\mathbf{r}_i\}\right)$.

We begin by considering square $L_A \times L_A$ subregions embedded within total systems of size $L\times L$ at $\rho=1/4$.  The second R\'enyi entropy $S_2$ for the free fermion state $\Psi_f^\mathrm{FS}$ on systems with $L=24,72,120$ as calculated via the correlation matrix technique \cite{Peschel03_JPhysA_36_L205, Peschel09_JPhysA_42_504003} is shown in Fig.~\ref{fig:Fig1}(d).  Plotting $S_2/L_A$ versus $\log L_A$ clearly reveals the multiplicative log violation.  We fit the $L=120$ data with $L_A$ between 4 and 36 to obtain an accurate linear fit $S_2/L_A = \kappa\log L_A + a$ with $\kappa=\kappa_W \equiv 0.2950(6)$ and $a=0.436(2)$.  The fitted value $\kappa_W$ is expected to be very close to that predicted by the Widom formula \cite{Barthel06_PRA_74_022329, Kim15_PRL_114_206402, Drut16_UnitaryGasEE_arXiv_1605.07085}.  The free fermion entropy for the gapped $d_{1,2}$ partons at $\nu=1$ is also shown in Fig.~\ref{fig:Fig1}(d); in this case, saturation to a boundary law is evident.

We now turn to Monte Carlo measurements of $S_2$.  As has become standard, we compute $S_2$ via the expectation value of the ``swap'' operator \cite{Hastings10_PRL_104_157201, Grover11_PRL_107_067202}:  $S_2=-\log[\mathrm{Tr}(\rho_A^2)] = -\log\langle\mathrm{SWAP}_A\rangle$.
(An alternative approach in the context of fermionic determinantal QMC was developed in Ref.~\cite{Grover13_DQMCRenyi_PRL_111_130402}; see also Refs.~\cite{Broecker14_RenyiDQMC_JStatMech_8_P08015, Trebst15_QMCRenyi_arXiv_1511.02878}.)
Importantly, we employ \cite{Note2} the mod/sign decomposition \cite{Grover11_PRL_107_067202} to compute the total R\'enyi entropy as a sum of two terms:  $S_2 = S_{2,\mathrm{total}} = S_{2,\mathrm{mod}} + S_{2,\mathrm{sign}}$ \footnote{$S_{2,\mathrm{mod}}$ is the entropy of the modulus of the wavefunction in the coordinate basis, while $S_{2,\mathrm{sign}}$ is the component of the entropy due to nontrivial signs (phases).  See the Supplemental Material for more details and discussion.}.  
We will argue that it is $S_{2,\mathrm{sign}}$ which is responsible for Eq.~\eqref{eq:LlogL} on long scales (cf.~Ref.~\cite{Grover11_PRL_107_067202}); hence, this approach allows us to glean more valuable long-distance information about $\kappa$ than what is contained in $S_{2,\mathrm{total}}$ alone.

\begin{figure*}
\centerline{
\hfill
\subfigure{\includegraphics[width=0.28\textwidth]{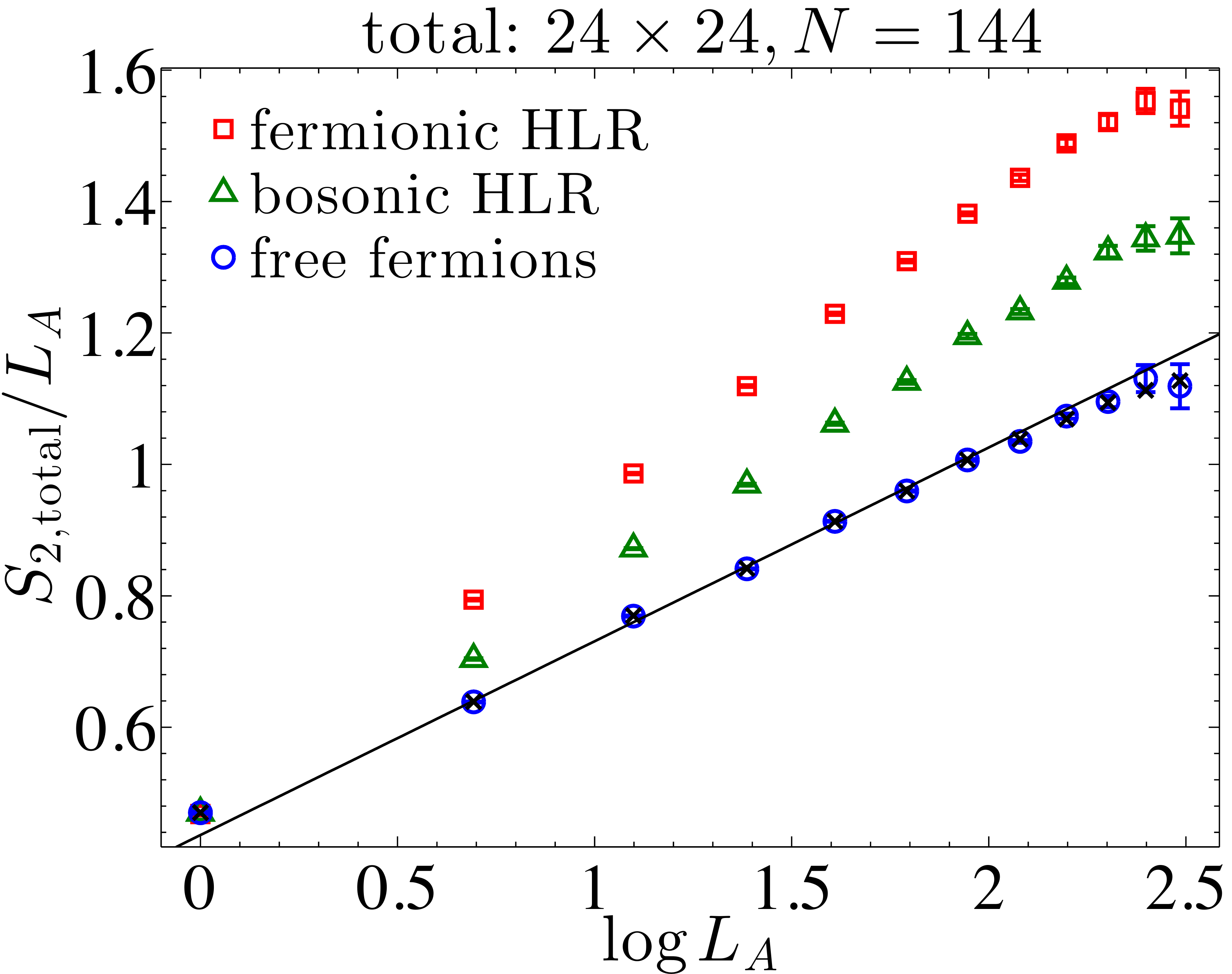}} \hfill
\subfigure{\includegraphics[width=0.28\textwidth]{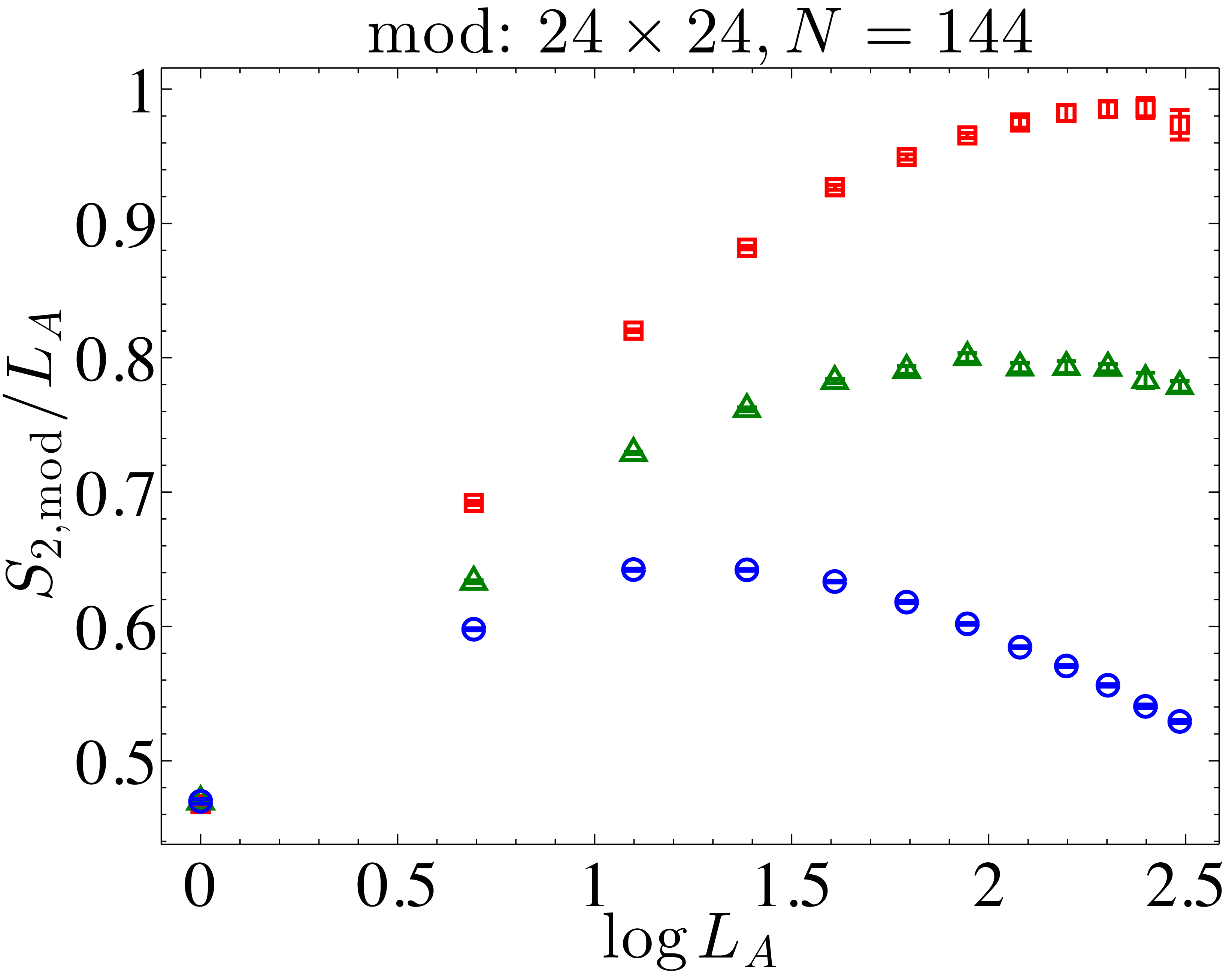}} \hfill
\subfigure{\includegraphics[width=0.28\textwidth]{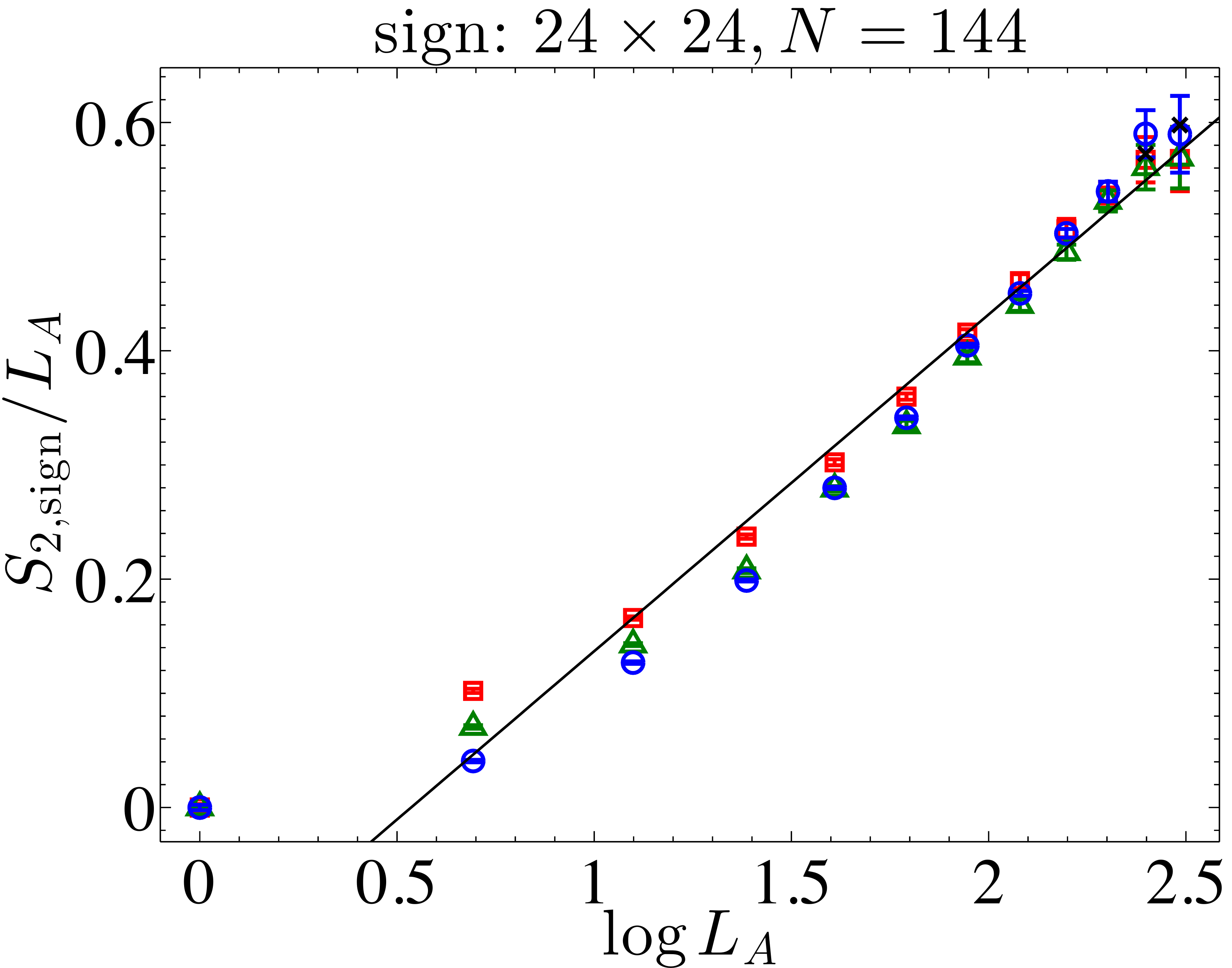}}
\hfill
}
\caption{Monte Carlo calculations of the total R\'enyi entropy (left panel) and the modulus (middle panel) and sign (right panel) components for the fermionic HLR, bosonic HLR, and free fermion wavefunctions on the $24\times24, N=144$ system with $L_A\times L_A$ subregions.  Here, and in Figs.~\ref{fig:Fig1}(d) and \ref{fig:2Dfs}, $L_A$ ranges from 1 to $L/2$.  The black ``$\times$'' symbols indicate the numerically exact $S_2$ values for free fermions \cite{Peschel03_JPhysA_36_L205, Peschel09_JPhysA_42_504003} (also in Figs.~\ref{fig:2Dfs} and \ref{fig:strips}), and the black lines indicate the Widom formula slope $\kappa_W$ from Fig.~\ref{fig:Fig1}(d).}
\label{fig:2D}
\end{figure*}

We show in Fig.~\ref{fig:2D} calculations of $S_{2,\mathrm{total}}$ (left panel), $S_{2,\mathrm{mod}}$ (middle panel), and $S_{2,\mathrm{sign}}$ (right panel) for both the fermionic and bosonic HLR wavefunctions, as well as for the free fermion wavefunction, on a $24\times24$ system with $N=144$ electrons.  As is evident in the left panel of Fig.~\ref{fig:2D}, the total entropy for the HLR wavefunctions indeed appears to have a slope $\kappa$ significantly enhanced over the free fermion/Widom value.
For example, fits to the fermionic HLR data indicate a $\kappa$ at least $60\%$ larger than that obtained by similar fits to the free fermion data.  We can thus corroborate the result of Ref.~\cite{Kim15_PRL_114_206402}:  For square subregions with $O(100)$ electrons, the HLR wavefunction appears to violate the Widom formula by nearly a factor of two.

However, a closer inspection of the contributions from the modulus and sign of the wavefunctions, as shown in the middle and right panels of Fig.~\ref{fig:2D}, reveals that this data is likely plagued by strong finite-size effects.  The dramatic increase in entanglement for the HLR wavefunctions is almost entirely due to contributions from $S_{2,\mathrm{mod}}$ on these sizes, while $S_{2,\mathrm{sign}}$ is remarkably nearly equal for all three wavefunctions.  However, $S_{2,\mathrm{mod}}$ displays eventual boundary law behavior (with quite large boundary law coefficients for the HLR wavefunctions).  On the other hand, it is clearly $S_{2,\mathrm{sign}}$ which is ultimately responsible for the long-distance $L_A\log L_A$ scaling behavior.  Hence, in order to make conclusions about $\kappa$ by analyzing only $S_{2,\mathrm{total}}$, one should be deep in a regime of $L_A$ where $S_{2,\mathrm{mod}}$ has saturated to a boundary law.

\begin{figure}[b]
\centerline{
\subfigure{\includegraphics[width=0.5\columnwidth]{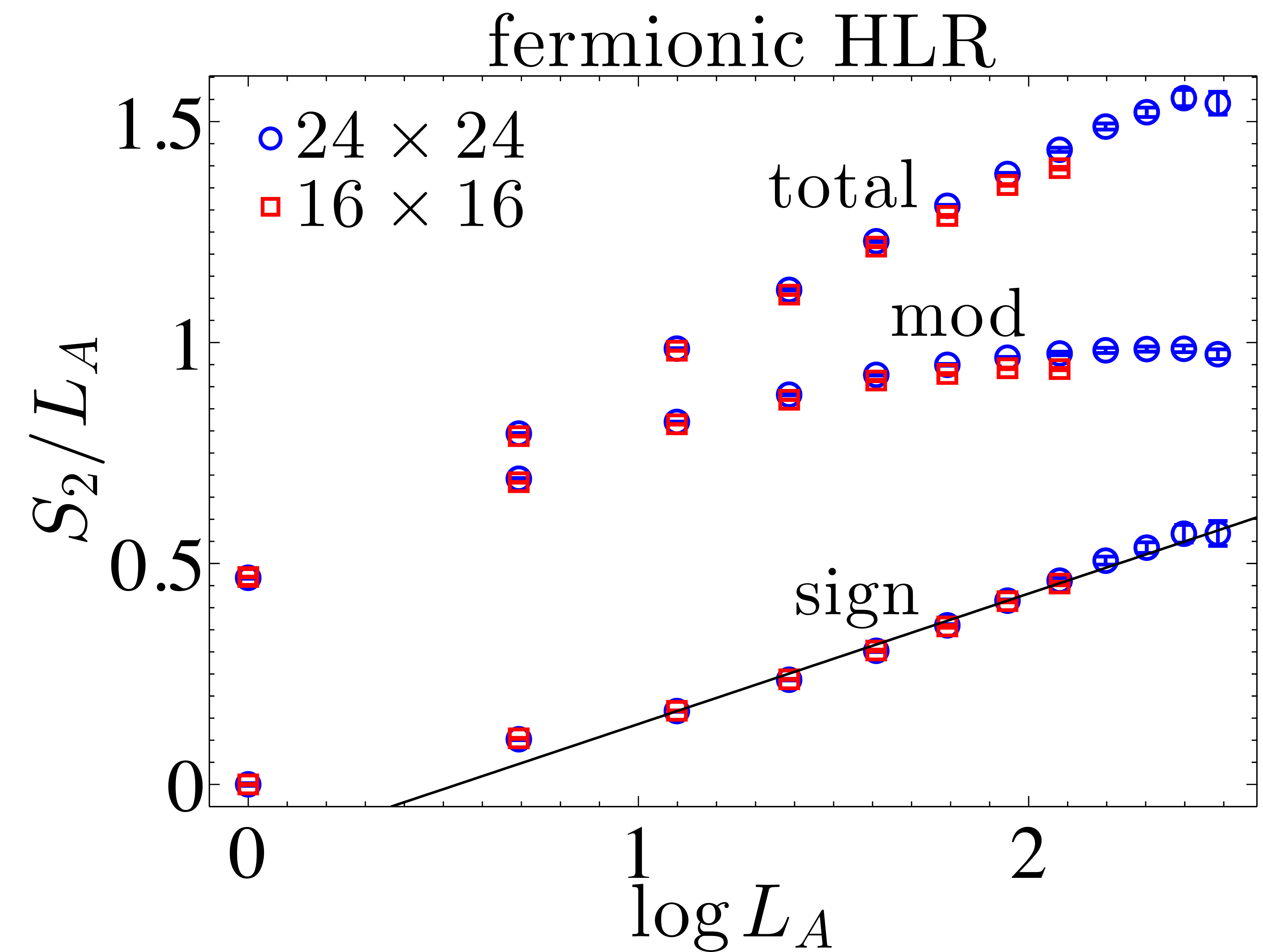}} \hspace{0.02in}
\subfigure{\includegraphics[width=0.49\columnwidth]{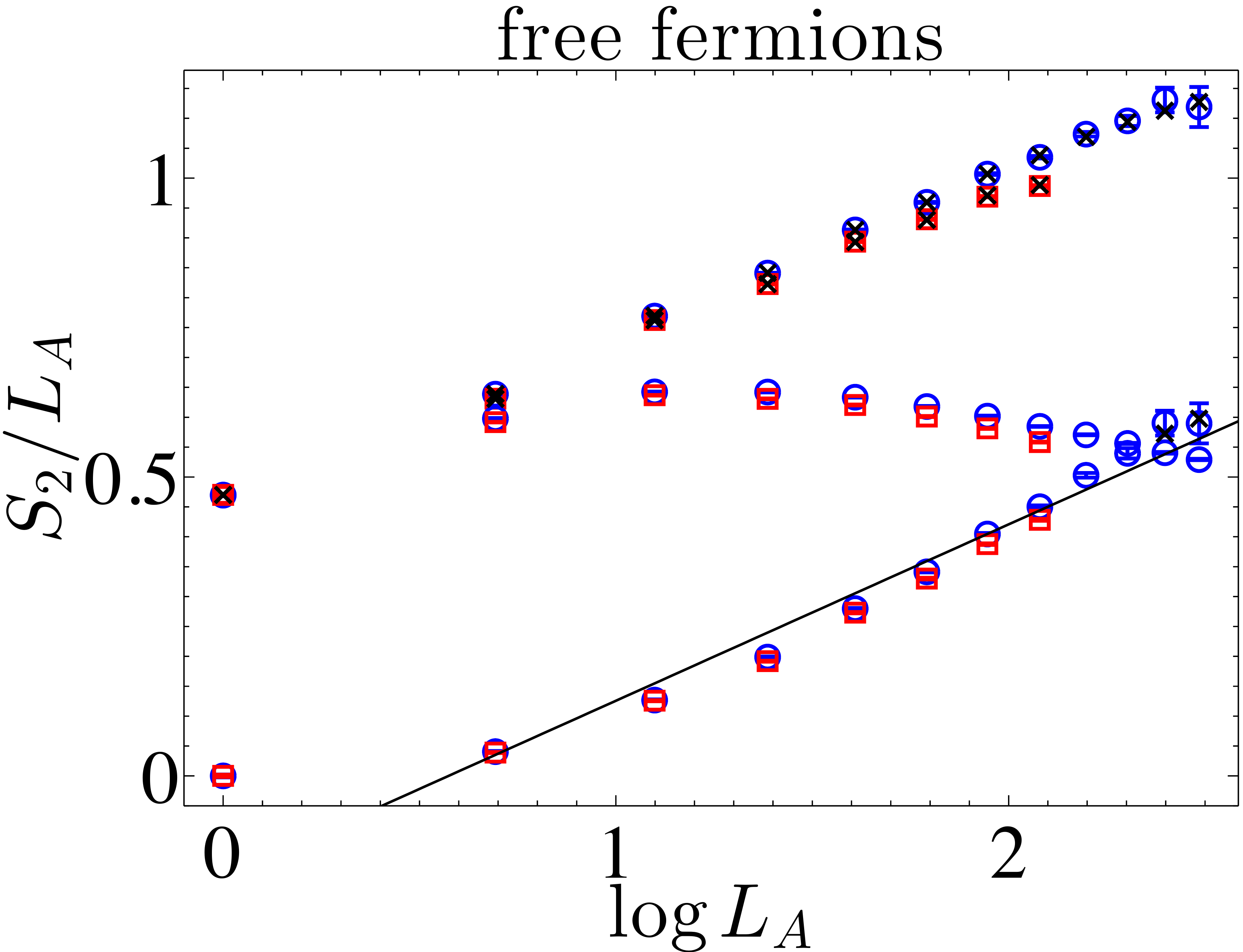}}
}
\caption{Fermionic HLR (left panel) and free fermion (right panel) data for $24\times24, N=144$ and $16\times16, N=64$ showing $S_{2,\mathrm{total}}$, $S_{2,\mathrm{mod}}$, and $S_{2,\mathrm{sign}}$ on the same axes.}
\label{fig:2Dfs}
\end{figure}

While for the HLR states we are not yet in such a regime on the $24\times24$, $N=144$ system \footnote{Even doubling $L$ (i.e., $48\times48$, $N=576$), which may or may not be sufficient, is already well out of current computational abilities (cf.~Refs.~\cite{McMinis13_PRB_87_081108, McMinis13_PhDThesis, Clark16_FracChernVMC_PRB_93_035125}).}, there are already telling indications in the $S_{2,\mathrm{sign}}$ data that these wavefunctions indeed do obey the Widom formula.  In the right panel of Fig.~\ref{fig:2D}, we show a line with slope $\kappa_W$ (intercept is arbitrary here and in Fig.~\ref{fig:2Dfs}).  For $L_A$ beyond just a couple lattice spacings, we see that $S_{2,\mathrm{sign}}$ very nearly obeys the Widom formula for all three wavefunctions, perhaps most accurately for the fermionic HLR state itself.
Finally, in Fig.~\ref{fig:2Dfs} we show an alternative view of the fermionic HLR (left panel) and free fermion (right panel) data from Fig.~\ref{fig:2D}, where we also include data from a smaller system:  $16\times16, N=64$.  As in the right panel of Fig.~\ref{fig:2D}, the black lines near the sign data indicate the Widom slope $\kappa_W$.  The following three points are now clear:  $(i)$ $S_{2,\mathrm{mod}}$ for $\Psi_\mathrm{HLR}^\mathrm{ferm}$ indeed saturates to a boundary law; $(ii)$ $S_{2,\mathrm{sign}}$ for $\Psi_\mathrm{HLR}^\mathrm{ferm}$ is well described by the Widom formula \footnote{Comparing the $16\times16$ and $24\times24$ data, $S_{2,\mathrm{sign}}/L_A$ appears to be well-converged in system size for $L_A=1-8$.}; and $(iii)$ the apparent Widom formula violation in $S_{2,\mathrm{total}}$ for $\Psi_\mathrm{HLR}^\mathrm{ferm}$ is mainly due to significant short-distance entanglement increase in the modulus of the wavefunction which results from strong correlations contained in the Jastrow-like factor $\left|\Psi_b^{(\nu=1/2)}\right|$
\footnote{Even though the pseudo-potential corresponding to the Jastrow factor $\left|\Psi_b^{(\nu=1/2)}\right|$ is very long-range ($\sim-\log r$), the entropy cannot increase stronger than boundary law since $S_{2,\mathrm{mod}} = S_{2,\mathrm{total}} - S_{2,\mathrm{sign}} \leq S_{2,\mathrm{total}}$ and $S_{2,\mathrm{total}}$ obeys a boundary law for the gapped state $\Psi_b^{(\nu=1/2)}$}.  Collectively, these three points suggest that the Widom formula will eventually be satisfied in the thermodynamic limit.

\begin{figure*}
\centerline{
\hfill
\subfigure{\includegraphics[width=0.28\textwidth]{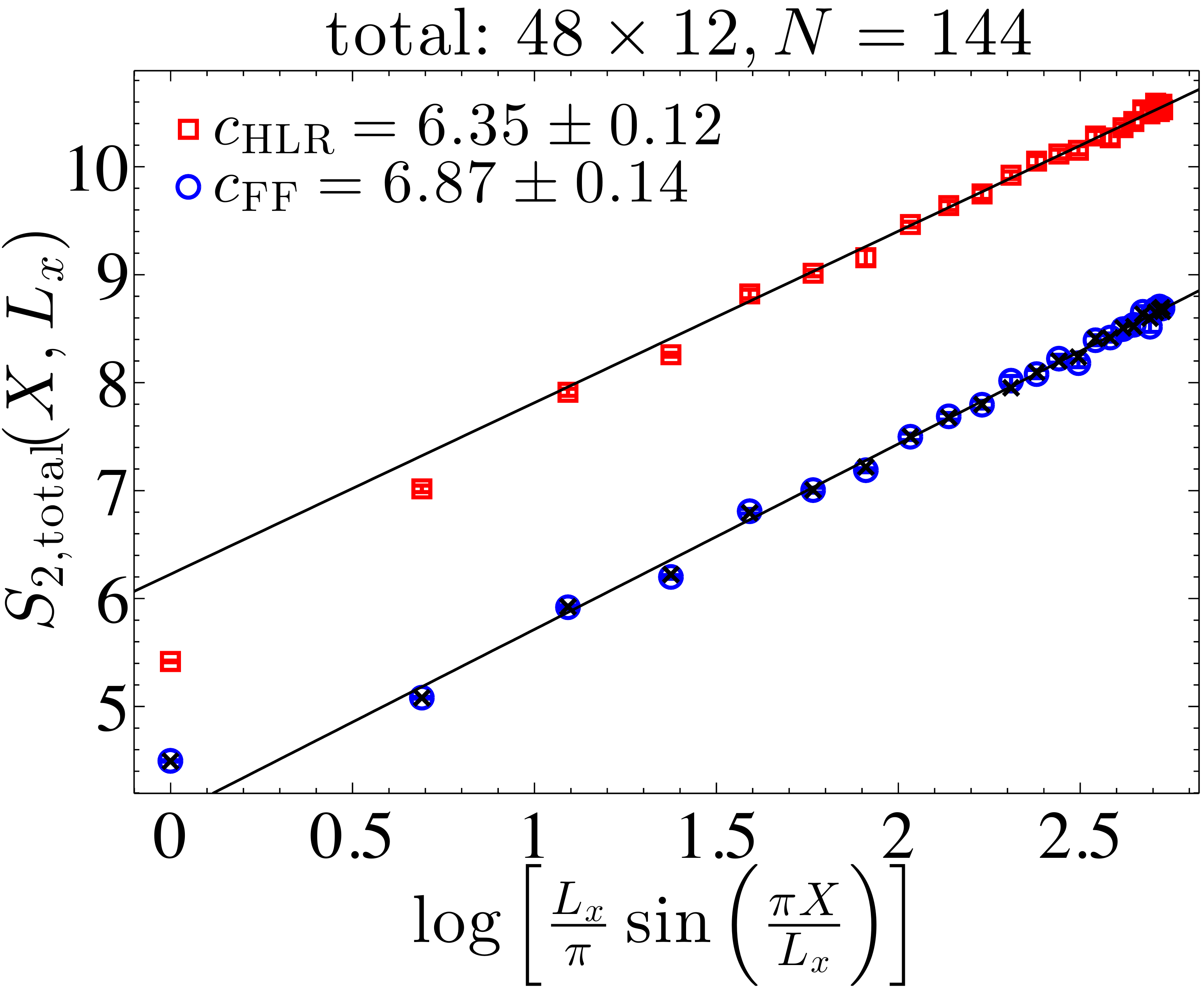}} \hfill
\llap{
	\raisebox{0.37in}{\hspace{-1in}\includegraphics[width=0.065\textwidth]{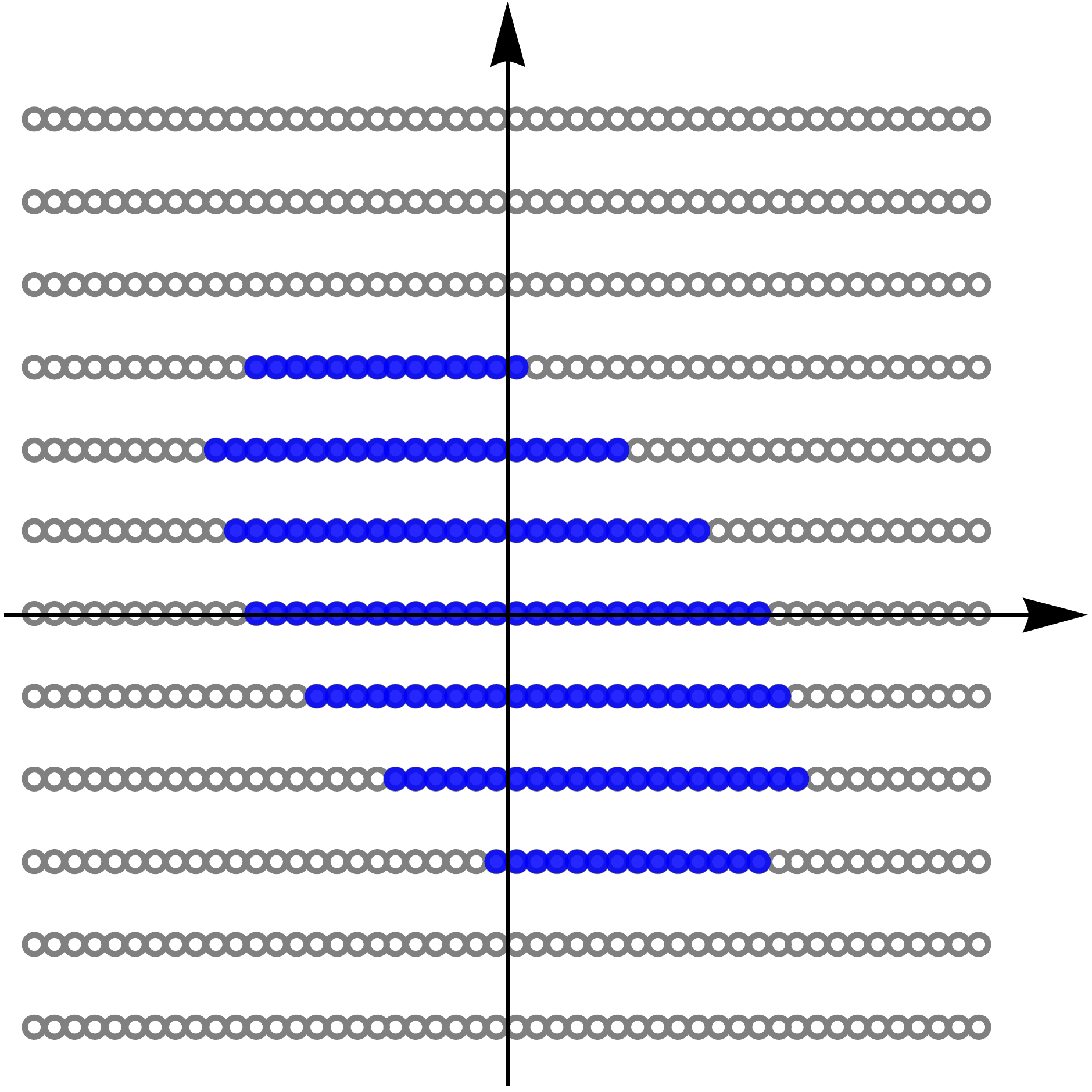}}
	\hspace{0.2in}
}
\subfigure{\includegraphics[width=0.28\textwidth]{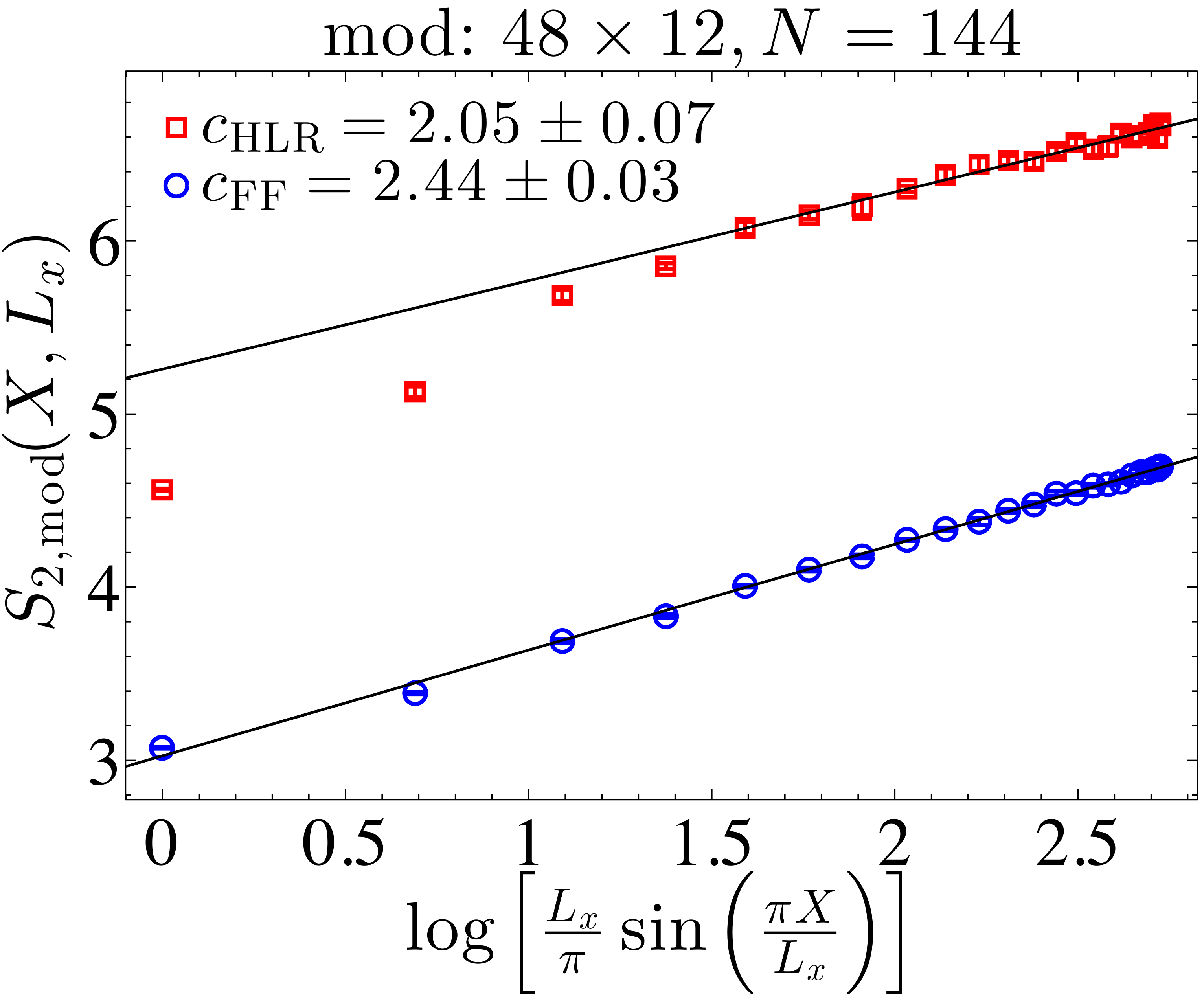}} \hfill
\subfigure{\includegraphics[width=0.28\textwidth]{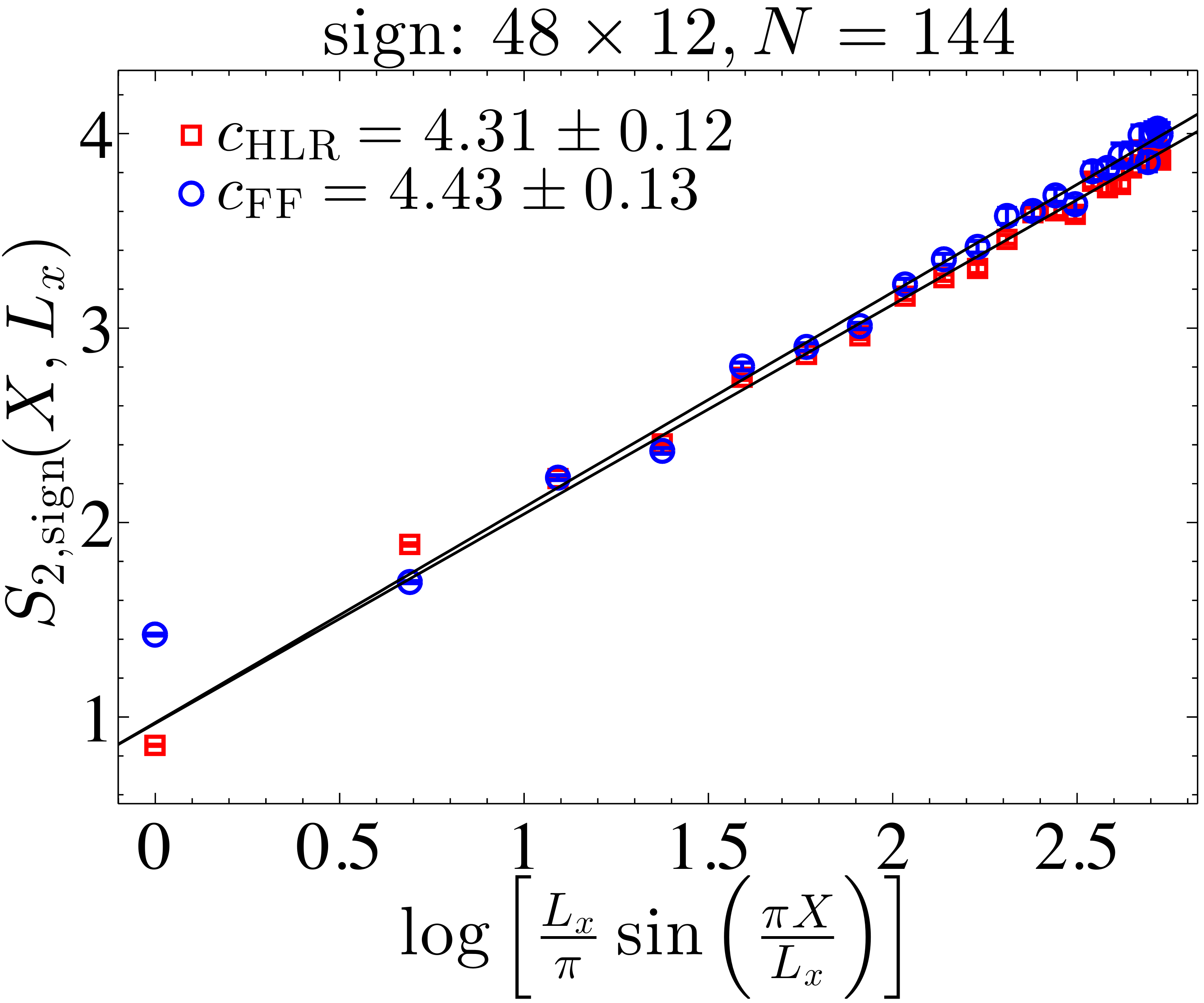}} \hfill
\hfill
}
\caption{From left to right, we show $S_{2,\mathrm{total}}$, $S_{2,\mathrm{mod}}$, and $S_{2,\mathrm{sign}}$ versus $\log\ell$ on a $48\times12, N=144$ system.  The free fermion (FF) state has $N_\mathrm{slices}=7$ [see inset in the left panel; cf.~Fig.~\ref{fig:Fig1}(a)].  The lines correspond to fits to Eq.~\eqref{eq:Cardy} with obtained values of $c$ given in the legends.}
\label{fig:strips}
\end{figure*}

We now further bolster our arguments that the fermionic HLR state obeys the Widom formula by considering $S_2$ scaling on \emph{strip} geometries.  That is, we take $X\times L_y$ subregions embedded within $L_x\times L_y$ systems and vary $X$.  In this case, for free fermions the Widom formula essentially reduces to the familiar quasi-1D form:
\begin{equation}
S_2(X,L_x) = \frac{c}{4}\log\left[\frac{L_x}{\pi}\sin\left(\frac{\pi X}{L_x}\right)\right] + A,
\label{eq:Cardy}
\end{equation}
where $c=N_\mathrm{slices}$ is simply the number of ``slices'' through which the quantized $k_y$ momenta pierce the Fermi surface, and we have used the familiar chord length $\ell$ inside the log \cite{Ju12_PRB_85_165121} (appropriate for $X$ comparable to $L_x$).  More generally, at least in the quasi-1D limit ($L_x\gg L_y$), $c$ is the \emph{central charge} \cite{Cardy04_JStatMech_P06002}, i.e., the number of (nonchiral) gapless modes present in the realized multimode Luttinger liquid \cite{Sheng08_PRB_78_054520, Sheng09_PRB_79_205112, Block11_PRL_106_046402, Mishmash11_PRB_84_245127, Jiang13_Nature_493_39}.

The narrowest nontrivial strip that we can consider has $L_y=4$ \cite{Note2} and $N_\mathrm{slices}=3$.  For free fermions, we thus expect an effective central charge $c=N_\mathrm{slices}=3$.  For the fermionic HLR state, on the other hand, we expect the Gutzwiller projection in Eq.~\eqref{eq:HLRwf} to \emph{remove} one gapless mode \cite{Sheng08_PRB_78_054520, Sheng09_PRB_79_205112, Block11_PRL_106_046402, Mishmash11_PRB_84_245127, Geraedts16_CFLDMRG_Science_352_197} giving $c=N_\mathrm{slices}-1=2$ (since $\Psi_b^{(\nu=1/2)}$ is fully gapped).  Indeed we can unambiguously confirm this prediction on a $48\times4, N=48$ system (see the Supplemental Material).

We have performed measurements on increasingly wide strips to approach the 2D limit.  By performing fits to the data using Eq.~\eqref{eq:Cardy}, we can extract the central charge associated with the total entropy, denoted $c_\mathrm{total}$, as well as contributions to the central charge from the mod and sign individually, denoted $c_\mathrm{mod}$ and $c_\mathrm{sign}$ (with $c_\mathrm{total}=c_\mathrm{mod}+c_\mathrm{sign}$).  Figure~\ref{fig:strips} shows an example of such data and the associated fits for $48\times12, N=144$.  This system has $N_\mathrm{slices}=7$, and indeed we find $c_\mathrm{total}\approx7$ for free fermions.  For the HLR state, $c_\mathrm{total}$ is reduced compared to free fermions and roughly consistent with $c\approx N_\mathrm{slices}-1$.

The middle and right panels of Fig.~\ref{fig:strips} again demonstrate that it is $S_{2,\mathrm{sign}}$ which is mainly responsible for the boundary law violation in these systems.  Remarkably, the fermionic HLR and free fermion $S_{2,\mathrm{sign}}$ results continue to track each other, both accurately following the scaling form Eq.~\eqref{eq:Cardy}.  On the other hand, $S_{2,\mathrm{mod}}$ grows relatively weakly with $\log\ell$ for both wavefunctions.  In fact, the main qualitative difference between the two states is simply a larger intercept $A$ in Eq.~\eqref{eq:Cardy} for the HLR state, which is coming entirely from the modulus of the wavefunction (consistent with Fig.~\ref{fig:2D}) and due to the presence of the $\Psi_{d_{1,2}}^{(\nu=1)}$. 
However, such physics is clearly distinct from that giving rise to the multiplicative log boundary law violation.

In the Supplemental Material, we present the entirety of our strip geometry study showing (in addition to Fig.~\ref{fig:strips}) simulations for $L_y=4,8,16$, and 20 with $L_x=48,48,36$, and 24, respectively, all at $\rho=1/4$.   As $L_y$ (and thus $N_\mathrm{slices}$) is increased, the \emph{scaling} of the entropy becomes concentrated in $c_\mathrm{sign}$ for both states (cf.~Fig.~\ref{fig:2D}) while $c_\mathrm{mod}$ remains of order one.  This itself constitutes a very interesting result---even for the free Fermi gas---which nicely elucidates the intimate relationship between sign structure and entanglement for these wavefunctions in the 2D limit.

All in all, we find no evidence that the HLR state violates the Widom formula in our strip geometry study, even in the total $S_2$ entropy itself.  That is, for the total entropy we have found $c_\mathrm{HLR}\approx c_\mathrm{FF}$ in all cases.  These results also put on firm footing the expression $c=N_\mathrm{slices}-1$ for the CFL used in the recent DMRG study of Ref.~\cite{Geraedts16_CFLDMRG_Science_352_197}.  It would be interesting to perform a similar analysis as we have in this work---for both types of subregion geometries---on the precise HLR wavefunction considered in Ref.~\cite{Kim15_PRL_114_206402}, and also on the interacting Fermi liquid wavefunctions considered in Ref.~\cite{McMinis13_PRB_87_081108} which were claimed to weakly violate the Widom formula.

While we have argued that our lattice HLR states have the same leading entanglement scaling as free fermions, it is interesting to think about which types of wavefunctions may actually violate the Widom formula
\footnote{For Gutzwiller-projected states containing multiple parton Fermi surfaces (see, e.g., Refs.~\cite{Motrunich05_PRB_72_045105, Motrunich07_PRB_75_235116, Jiang13_Nature_493_39}), we conjecture an effective Widom formula in which the total prefactor of $L_A \log L_A$ is given by the sum of the $\kappa_W$ prefactors for each Fermi surface individually.}.
On this note, we have also considered a wavefunction in the form of Eq.~\eqref{eq:HLRwf} but with $\Psi_b^{(\nu=1/2)}\to\Psi_b^{(\nu=1/2)}/|\Psi_b^{(\nu=1/2)}|$, i.e., a wavefunction with sign structure given by $\Psi_\mathrm{HLR}^\mathrm{ferm}$ but amplitudes given by $\Psi_f^\mathrm{FS}$.  Such wavefunctions basically model attachment of flux at the mean-field level---as opposed to attachment of vortices in Eq.~\eqref{eq:HLRwf}---and are known to have various deficiencies \cite{Jain89_PRL_63_199, Jain07_CFbook}.  Interestingly, we find that $S_{2,\mathrm{sign}}$ for this wavefunction, grows extremely quickly with $L_A$,
and the full wavefunction may possibly have a scaling different from the Widom formula.  We leave further investigation of this result for future work.  Finally, Gutzwiller projection---as employed here and, for example, in the spin liquid states in Ref.~\cite{Grover11_PRL_107_067202}---is known to only capture gauge fluctuations in a partial way \cite{Tay11_PRB_83_235122}.  Remedying this problem and subsequently studying the long-distance entanglement properties of such wavefunctions constitutes an exciting and challenging future direction.

\acknowledgments
We gratefully acknowledge Sarang Gopalakrishnan, Hsin-Hua Lai, Max Metlitski, David Mross, Mike Mulligan, Sri Raghu, and Ashvin Vishwanath for valuable discussions.  R.V.M. would especially like to thank Jim Garrison for explaining the second Monte Carlo scheme described in the Supplemental Material.  This work was supported by the NSF through grant DMR-1206096 (O.I.M.); the Caltech Institute for Quantum Information and Matter, an NSF Physics Frontiers Center with support of the Gordon and Betty Moore Foundation; and the Walter Burke Institute for Theoretical Physics at Caltech.  This work used the Extreme Science and Engineering Discovery Environment (XSEDE), which is supported by National Science Foundation grant number ACI-1053575.

\bibliography{HLR_EE}



\section{\underline{\uppercase{Supplemental Material}}}

\subsection{Details of the projected wavefunctions}

The orbitals for the Slater determinants $\Psi_{a=d_1,d_2,f}$ used in defining $\Psi_\mathrm{HLR}^\mathrm{ferm}$ and $\Psi_\mathrm{HLR}^\mathrm{bos}$ are obtained by diagonalizing mean-field hopping Hamiltonians of the form
\begin{eqnarray}
\hspace{-0.15in}
H_\mathrm{MF} = -\sum_{\mathbf{r}=(r_x,r_y)}&\,&\hspace{-0.15in}\biggl[t_{\hat{x}} e^{-i2r_y\phi}a_\mathbf{r}^\dagger a_{\mathbf{r}+\hat{x}} + t_{\hat{y}} a_\mathbf{r}^\dagger a_{\mathbf{r}+\hat{y}} \nonumber \\ &+& t_{\hat{x}+\hat{y}} e^{-i(2r_y+1)\phi}a_\mathbf{r}^\dagger a_{\mathbf{r}+\hat{x}+\hat{y}} + \hc\biggr]
\label{eq:MF}
\end{eqnarray}
and filling the lowest $N$ states.  Throughout, we use a ``squarized'' version of the triangular lattice as shown in Fig.~\ref{fig:Fig1}(c) with $r_x=0,1,\dots,L_x-1$ and $r_y=0,1,\dots,L_y-1$.  Equation~\eqref{eq:MF} corresponds to a Landau-like gauge giving uniform flux $\phi$ through each triangle.

At $\rho=1/4$, we take $\phi=\pi/4$ for $d_{1,2}$ corresponding to $\nu=1$ [see the filled nearly flat band in Fig.~\ref{fig:Fig1}(b); there, $M=(\pi,\pi/4)$ and $X=(\pi,0)$].  The magnetic unit cell for $d_{1,2}$ thus consists of four sites along a line in the $y$ direction, which is very natural for our torus geometry.  For the $f$ partons, $\phi=0$ [see the sharp Fermi surface in Fig.~\ref{fig:Fig1}(a)].

We choose completely isotropic hopping patterns $t_{\hat{x}}=t_{\hat{y}}=t_{\hat{x}+\hat{y}}=1$ for all partons except for the $48\times12, N=144$ system in Fig.~\ref{fig:strips} of the main text, where we take $t_{\hat{x}}=t_{\hat{y}}=1$ and $t_{\hat{x}+\hat{y}}=1.01$ for the $f$ partons to avoid degeneracies at the Fermi energy.  The boundary conditions are taken to be periodic in the $y$ direction for all partons and antiperiodic in the $x$ direction for all partons except $d_2$; this produces a wavefunction with periodic boundary conditions in both directions.

\subsection{Details of the Monte Carlo simulations}

Given a wavefunction in coordinate space, $\phi(\alpha)$, the expectation value of the swap operator is given by \cite{Hastings10_PRL_104_157201, Grover11_PRL_107_067202}
\begin{equation}
\langle\mathrm{SWAP}_A\rangle = \sum_{\alpha_1,\alpha_2}\frac{|\phi(\alpha_1)|^2}{\mathcal{N}}\frac{|\phi(\alpha_2)|^2}{\mathcal{N}}\left[\frac{\phi(\beta_1)\phi(\beta_2)}{\phi(\alpha_1)\phi(\alpha_2)}\right].
\end{equation}
Here $\alpha_1=(a_1,b_1)$ and $\alpha_2=(a_2,b_2)$ are configurations of the two copies 1 and 2 ($a$ refers to degrees of freedom in subregion $A$, whereas $b$ refers to degrees of freedom in the complement of $A$), while $\beta_1=(a_2,b_1)$ and $\beta_2=(a_1,b_2)$ are the swapped configurations, and $\mathcal{N}=\sum_{\alpha}|\phi(\alpha)|^2$ is the wavefunction normalization.

The mod/sign decomposition \cite{Grover11_PRL_107_067202} is given by
\begin{eqnarray}
\label{eq:modsign}
\langle\mathrm{SWAP}_A\rangle &=& \langle\mathrm{SWAP}_{A,\mathrm{mod}}\rangle \langle\mathrm{SWAP}_{A,\mathrm{sign}}\rangle \\
\langle\mathrm{SWAP}_{A,\mathrm{mod}}\rangle &=& \sum_{\alpha_1,\alpha_2}\frac{|\phi(\alpha_1)|^2}{\mathcal{N}}\frac{|\phi(\alpha_2)|^2}{\mathcal{N}}\left|\frac{\phi(\beta_1)\phi(\beta_2)}{\phi(\alpha_1)\phi(\alpha_2)}\right|, \nonumber \\
\langle\mathrm{SWAP}_{A,\mathrm{sign}}\rangle &=& \sum_{\alpha_1,\alpha_2}\frac{|\phi(\alpha_1)\phi(\alpha_2)\phi(\beta_1)\phi(\beta_2)|}{\mathcal{M}}e^{i\theta(\alpha_1,\alpha_2)}, \nonumber
\end{eqnarray}
where $\theta(\alpha_1,\alpha_2)=\arg[\phi^*(\alpha_1)\phi^*(\alpha_2)\phi(\beta_1)\phi(\beta_2)]$ and $\mathcal{M}=\sum_{\alpha_1,\alpha_2}|\phi(\alpha_1)\phi(\alpha_2)\phi(\beta_1)\phi(\beta_2)|$.  Hence, $S_2=S_{2,\mathrm{total}}=S_{2,\mathrm{mod}}+S_{2,\mathrm{sign}}$, with $S_{2,\mathrm{mod/sign}}=-\log\langle\mathrm{SWAP}_{A,\mathrm{mod/sign}}\rangle$.

Since $\langle\mathrm{SWAP}_{A,\mathrm{mod}}\rangle$ is the swap operator evaluated for the modulus of the wavefunction in this basis, i.e., $|\phi(\alpha)|$, $S_{2,\mathrm{mod}}$ is the entropy of the wavefunction $|\phi(\alpha)|$.  On the other hand, $S_{2,\mathrm{sign}}$ can be interpreted as the component of the entropy \emph{as a result of} nontrivial signs in the wavefunction:  For a positive wavefunction, most notably $|\phi(\alpha)|$, $S_{2,\mathrm{sign}}=0$ vanishes identically.  [Note that $S_{2,\mathrm{sign}}$ is \emph{not} simply the entropy obtained after taking the sign of the wavefunction $\phi(\alpha)\to\phi(\alpha)/|\phi(\alpha)|$; it depends in a specific way on the amplitudes as well.]

For systems with a globally conserved $U(1)$ symmetry---such as particle number conservation present in the wavefunctions in this work---it affords to be smart when performing the Monte Carlo walks in Eq.~\eqref{eq:modsign}:  Only configurations for which the total subregion occupations $N_A$ in the two copies are identical [i.e., $N_A(\alpha_1)=N_A(\alpha_2)=n_A$] give nonzero contributions.  [If $N_A(\alpha_1)\neq N_A(\alpha_2)$, then $\phi(\beta_1)=\phi(\beta_2)=0$.]  We have implemented two schemes for sampling ${\langle}\mathrm{SWAP}_A{\rangle}$, both of which allow the mod/sign factorization described above and which take advantage of the global particle number conservation (see also, e.g., Refs.~\cite{McMinis13_PRB_87_081108, McMinis13_PhDThesis, Clark16_FracChernVMC_PRB_93_035125} for similar schemes).

The first is the ``particle number trick'' explained in Ref.~\cite{Kim15_PRL_114_206402}, which we briefly review.  In this case, we decompose the final measurement as a sum over the possible subregion particle occupation numbers $n_A$:
\begin{equation}
{\langle}\mathrm{SWAP}_A{\rangle} = \sum_{n_A} \left(P_{n_A}\right)^2\langle\mathrm{SWAP}_{A,\mathrm{mod}}\rangle_{n_A} \langle\mathrm{SWAP}_{A,\mathrm{sign}}\rangle_{n_A}
\label{eq:sectored}
\end{equation}
Here, $\langle\mathrm{SWAP}_{A,\mathrm{mod/sign}}\rangle_{n_A}$ are the mod/sign measurements restricted to the subspace with $n_A$ particles in subregion $A$ for both copies.  [Formally, one just replaces all sums in the expressions in Eq.~\eqref{eq:modsign}---including those in the normalizations $\mathcal{N}$ and $\mathcal{M}$---with sums over the restricted subspace:  $\sum_{\alpha_1,\alpha_2}\to\sum_{\alpha_1,\alpha_2\in n_A}\equiv\sum_{\alpha_1,\alpha_2}\delta_{N_A(\alpha_1), n_A}\delta_{N_A(\alpha_2),n_A}$.]  The quantities $P_{n_A}$ are simply the probabilities of finding $n_A$ particles in subregion $A$ for a single copy of the wavefunction,
\begin{equation}
P_{n_A} = \sum_{\alpha\in n_A}\frac{|\phi(\alpha)|^2}{\mathcal{N}} = \sum_\alpha\frac{|\phi(\alpha)|^2}{\mathcal{N}}\delta_{N_A(\alpha), n_A},
\end{equation}
and are obtainable in a straightforward single-copy simulation.  In this scheme, we run separate swap simulations for each $n_A$ and compile the results according to Eq.~\eqref{eq:sectored}.  Note that $\langle\mathrm{SWAP}_{A,\mathrm{mod}}\rangle$ in Eq.~\eqref{eq:modsign} can be computed by performing the sum in Eq.~\eqref{eq:sectored} with $\langle\mathrm{SWAP}_{A,\mathrm{sign}}\rangle_{n_A}=1$; thus, the scheme readily gives both $S_{2,\mathrm{mod}}$ and $S_{2,\mathrm{sign}}=S_{2,\mathrm{total}}-S_{2,\mathrm{mod}}$.

The second method is similar to the original decomposition in Eq.~\eqref{eq:modsign}, except that for both the mod and sign walks we only consider in our move scheme ``swappable'' configurations, i.e., those with $N_A(\alpha_1)=N_A(\alpha_2)=n_A$, but we allow $n_A$ to fluctuate throughout the simulation.  Since the summands in the expressions in Eq.~\eqref{eq:modsign} are both proportional to $|\phi(\beta_1)\phi(\beta_2)|$, it is legitimate to replace the sums in the numerators with sums over only the swappable configurations:  $\sum_{\alpha_1,\alpha_2}\to\sum'_{\alpha_1,\alpha_2}\equiv\sum_{n_A}\sum_{\alpha_1,\alpha_2\in n_A}$.  Furthermore, since for the sign walk it is the \emph{weights} which contain $|\phi(\beta_1)\phi(\beta_2)|$, this replacement can also be performed in the expression for $\mathcal{M}$.  The final expression for $\langle\mathrm{SWAP}_{A,\mathrm{sign}}\rangle$ that we use for our simulations is thus given by that in Eq.~\eqref{eq:modsign} with $\sum_{\alpha_1,\alpha_2}\to\sum'_{\alpha_1,\alpha_2}$.  The mod case, on the other hand, requires a bit more care since it is now the \emph{measurements} which contain $|\phi(\beta_1)\phi(\beta_2)|$, so that the normalization $\mathcal{N}$ still contains an unrestricted sum over all configurations.  This is easily remedied with a small amount of algebra to give
\begin{equation}
\langle\mathrm{SWAP}_{A,\mathrm{mod}}\rangle = \left[\sum_{n_A}\left(P_{n_A}\right)^2\right]\langle\mathrm{SWAP}_{A,\mathrm{mod}}\rangle',
\end{equation}
where the first factor in brackets is the overall probability that the two-copy system is swappable, and
\begin{equation}
\langle\mathrm{SWAP}_{A,\mathrm{mod}}\rangle' = \sideset{}{'}\sum_{\alpha_1,\alpha_2}\frac{|\phi(\alpha_1)|^2|\phi(\alpha_2)|^2}{\mathcal{N}'}\left|\frac{\phi(\beta_1)\phi(\beta_2)}{\phi(\alpha_1)\phi(\alpha_2)}\right|
\end{equation}
is the mod calculation that we perform only over the swappable subspace [with $\mathcal{N}'=\sum_{\alpha_1,\alpha_2}'|\phi(\alpha_1)|^2|\phi(\alpha_2)|^2$].  In this scheme, since we are explicitly enforcing that the visited configurations are swappable, care must be taken to maintain detailed balance when the total subregion occupation number changes in a proposed move.  This consideration is valid for both the mod and sign walks.

\begin{figure}[b]
\centerline{
\subfigure{\includegraphics[height=0.5\columnwidth]{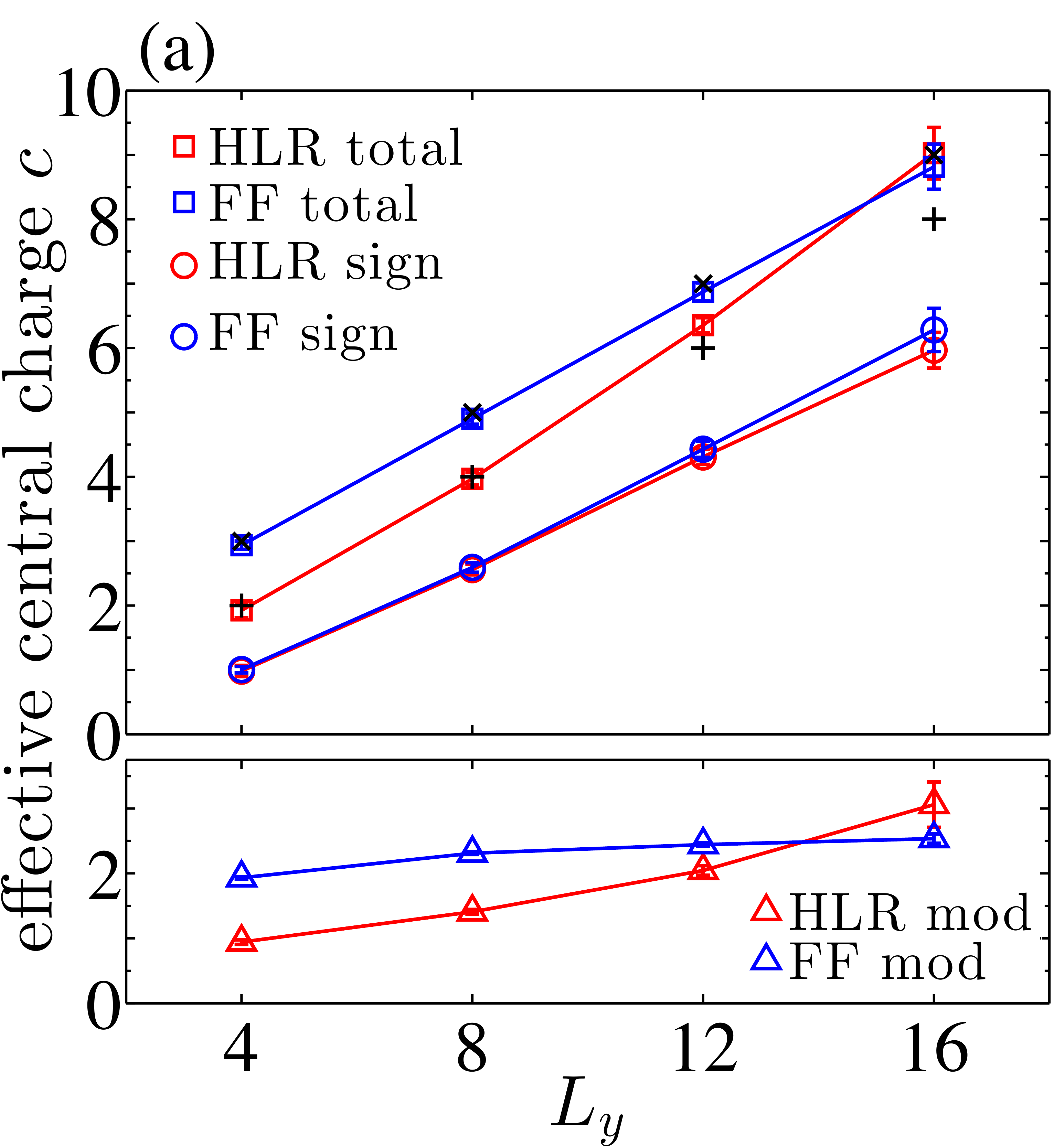}} \hspace{0.05in}
\subfigure{\includegraphics[height=0.5\columnwidth]{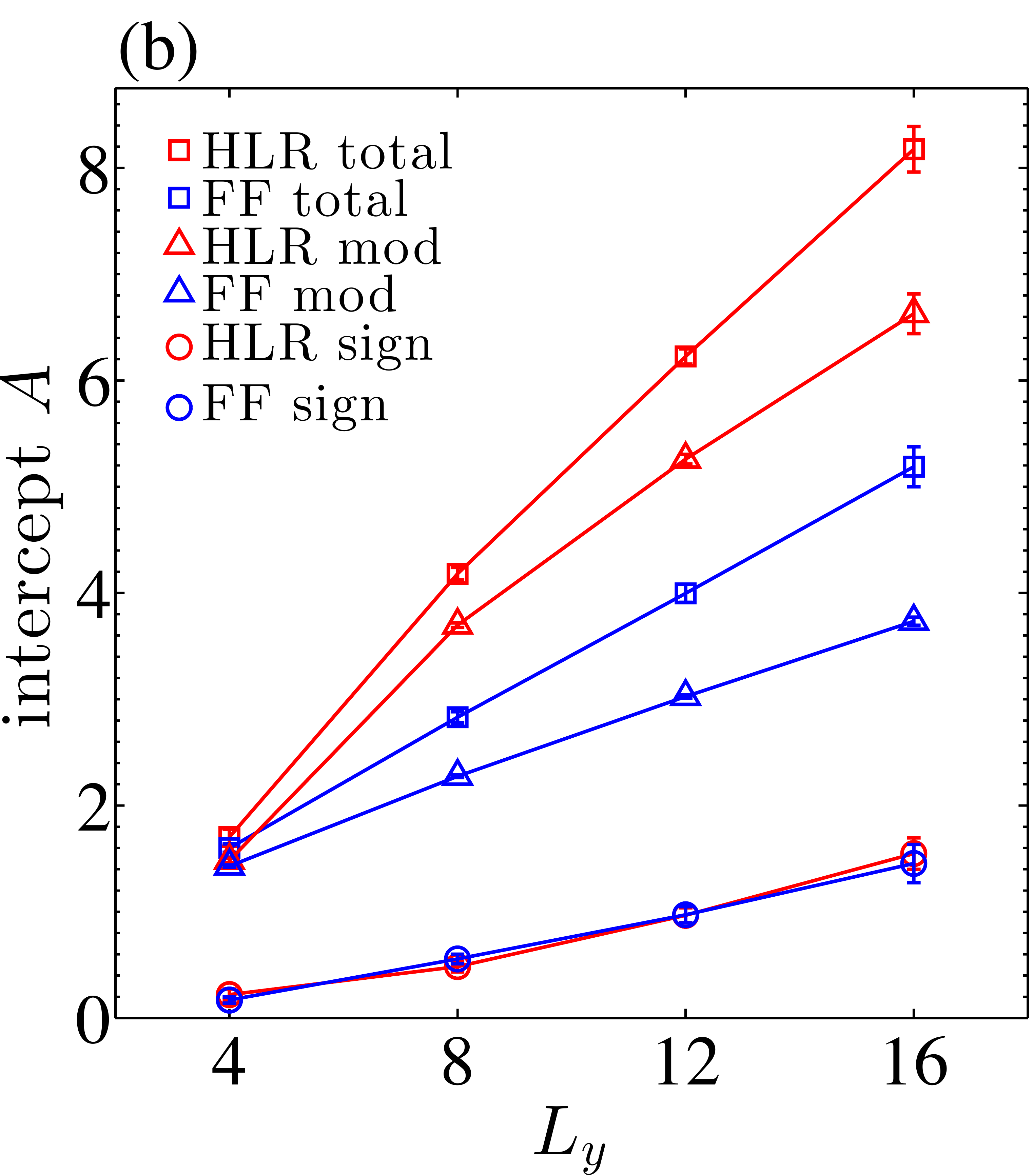}}
}
\caption{(a) Central charge $c$ and (b) intercept $A$ fit parameters versus $L_y$ for the fermionic HLR and free fermion (FF) wavefunctions on the strip geometry (lines are a guide to the eye).  In (a), ``$+$'' and ``$\times$''  symbols mark $N_\mathrm{slices}-1$ and $N_\mathrm{slices}$.}
\label{fig:fits}
\end{figure}

\begin{figure*}
\centerline{
\hfill
\subfigure{\includegraphics[width=0.28\textwidth]{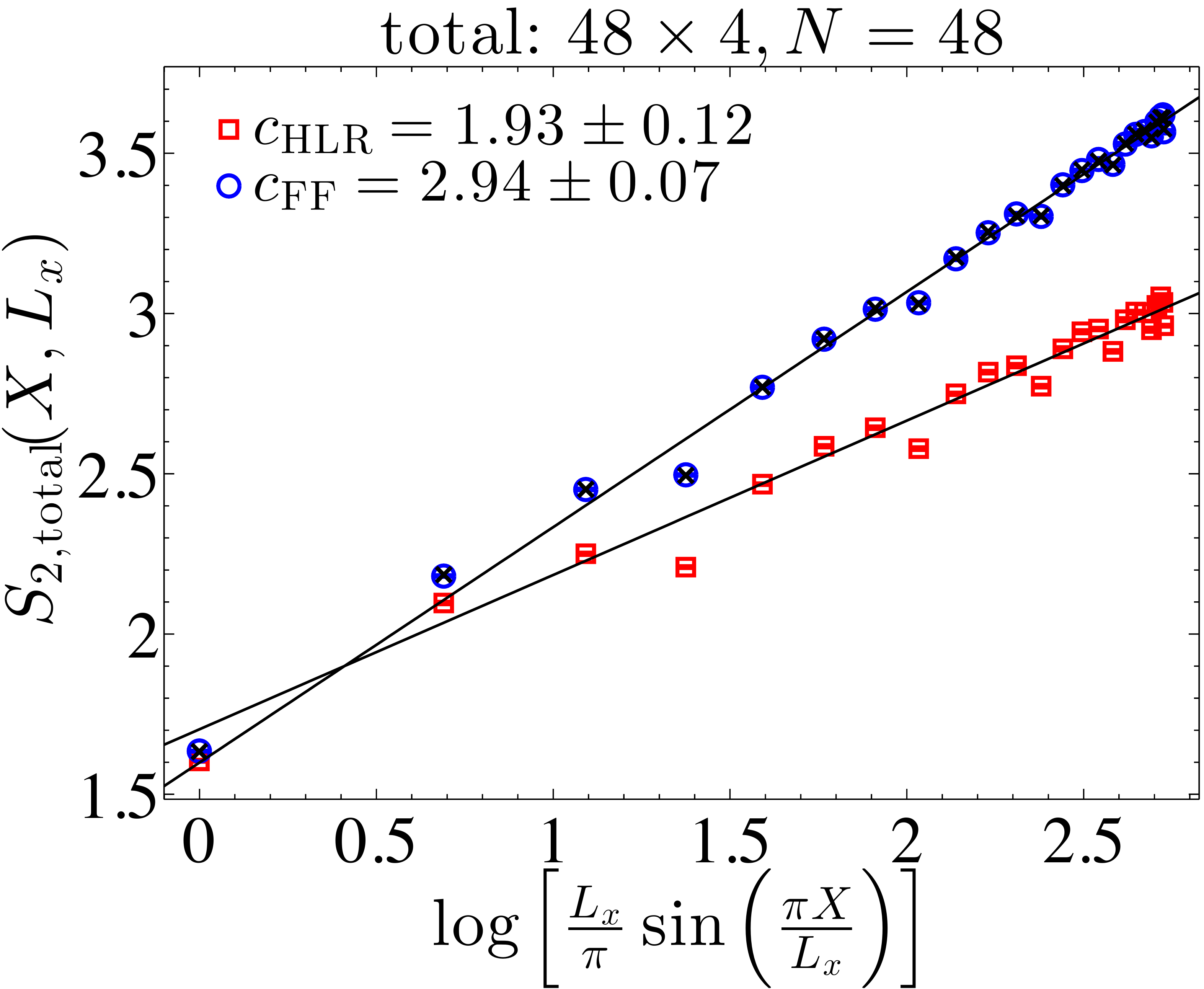}} \hfill
\subfigure{\includegraphics[width=0.28\textwidth]{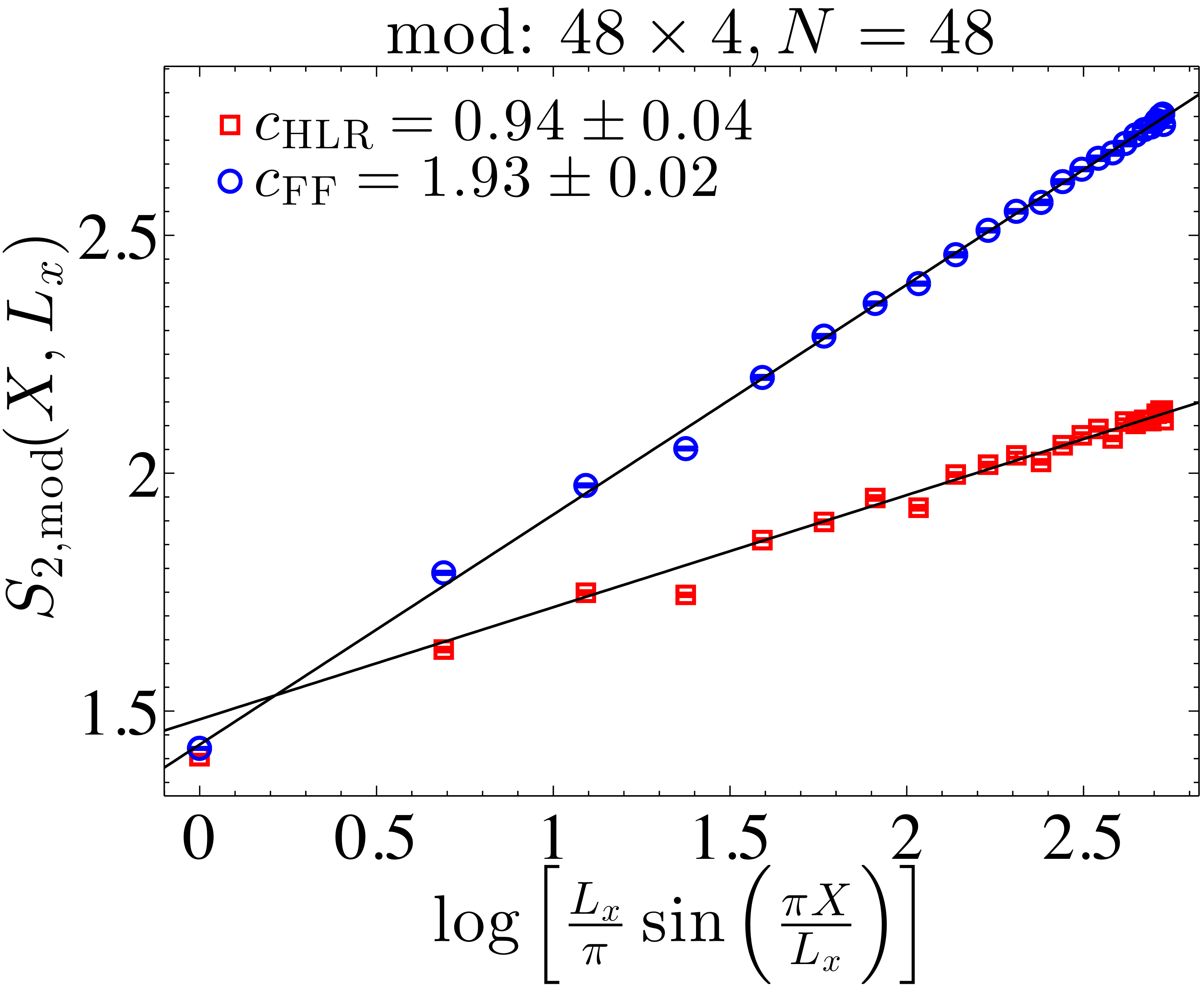}} \hfill
\subfigure{\includegraphics[width=0.28\textwidth]{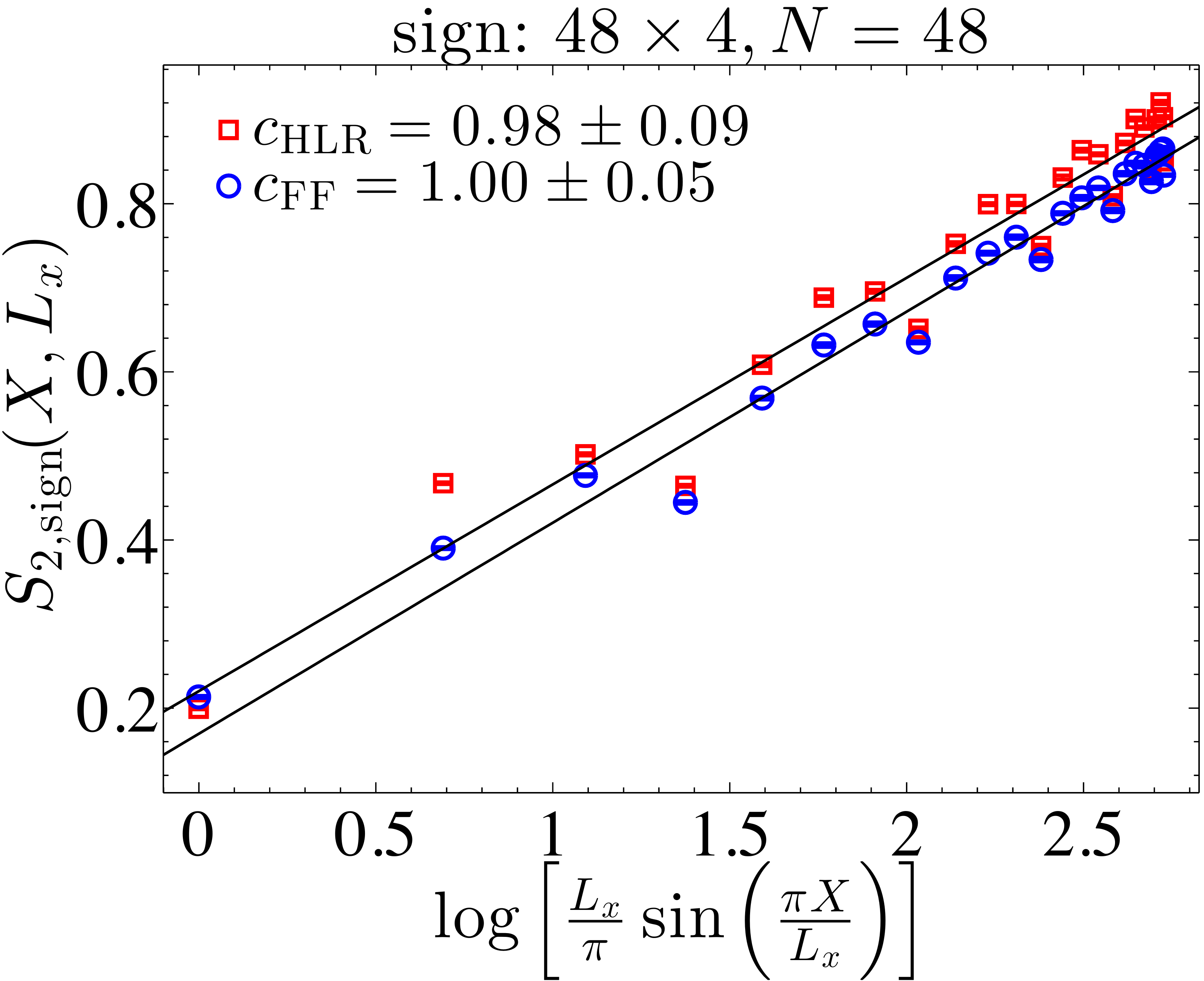}} \hfill
\hfill
}
\centerline{
\hfill
\subfigure{\includegraphics[width=0.28\textwidth]{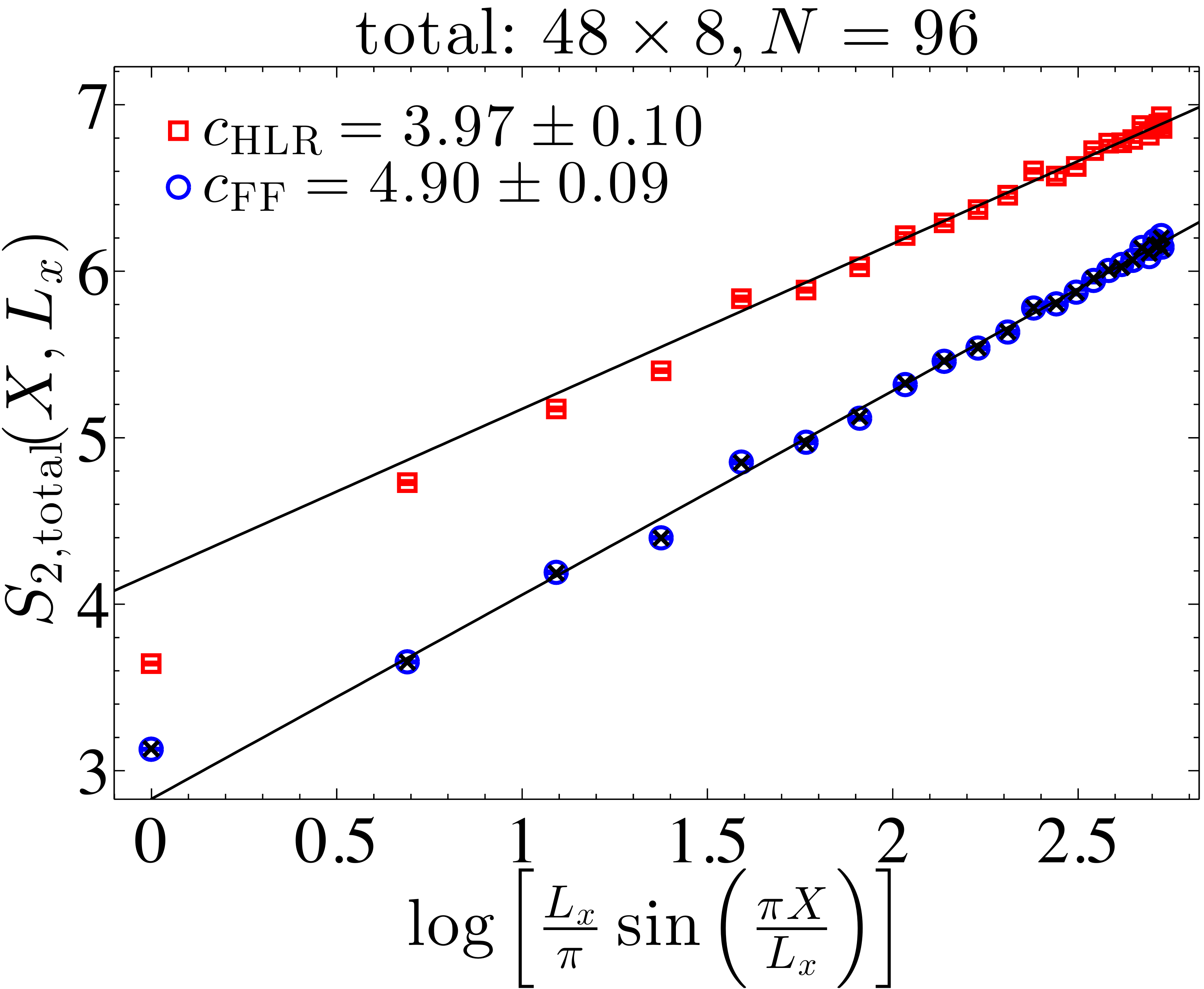}} \hfill
\subfigure{\includegraphics[width=0.28\textwidth]{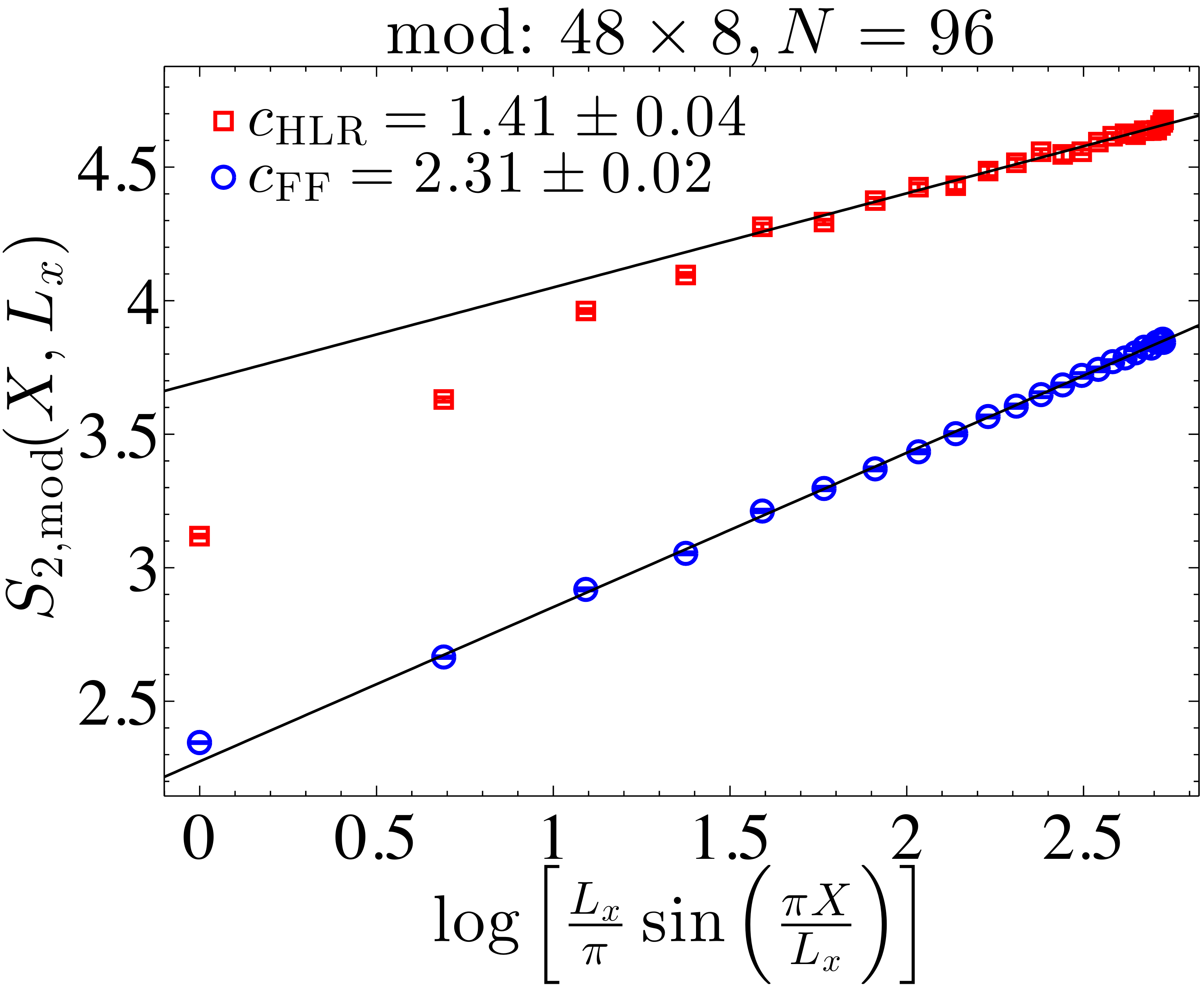}} \hfill
\subfigure{\includegraphics[width=0.28\textwidth]{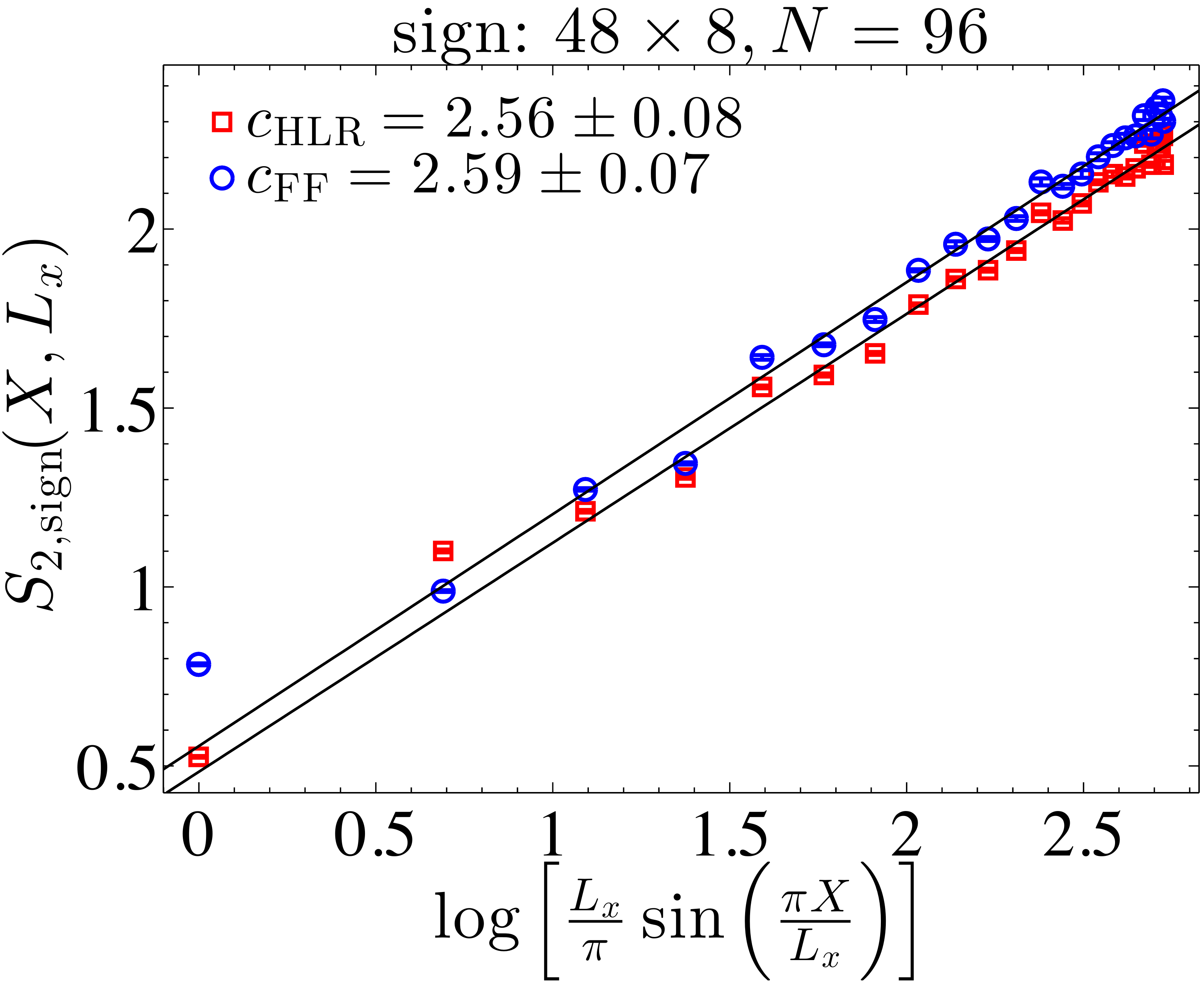}} \hfill
\hfill
}
\centerline{
\hfill
\subfigure{\includegraphics[width=0.28\textwidth]{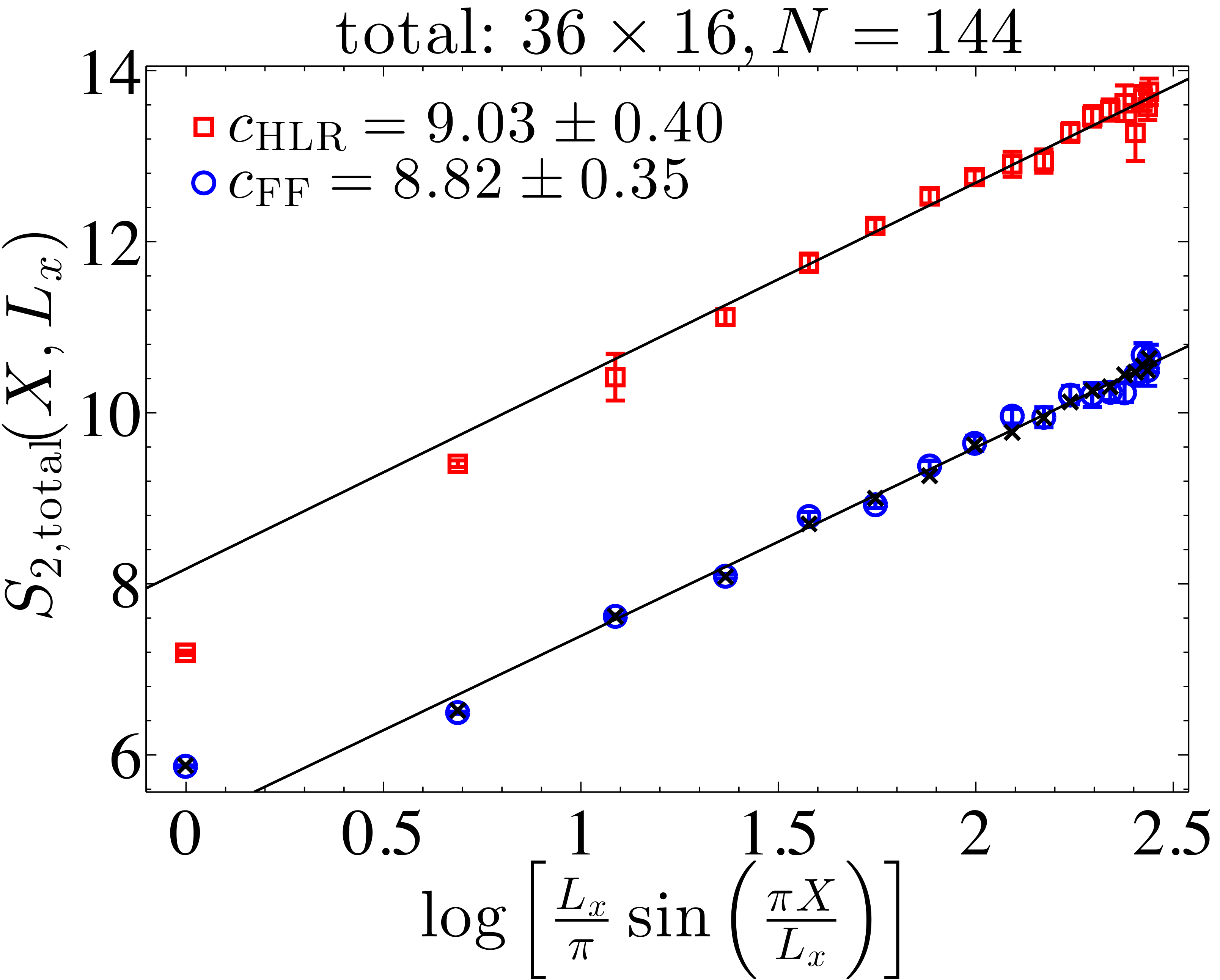}} \hfill
\subfigure{\includegraphics[width=0.28\textwidth]{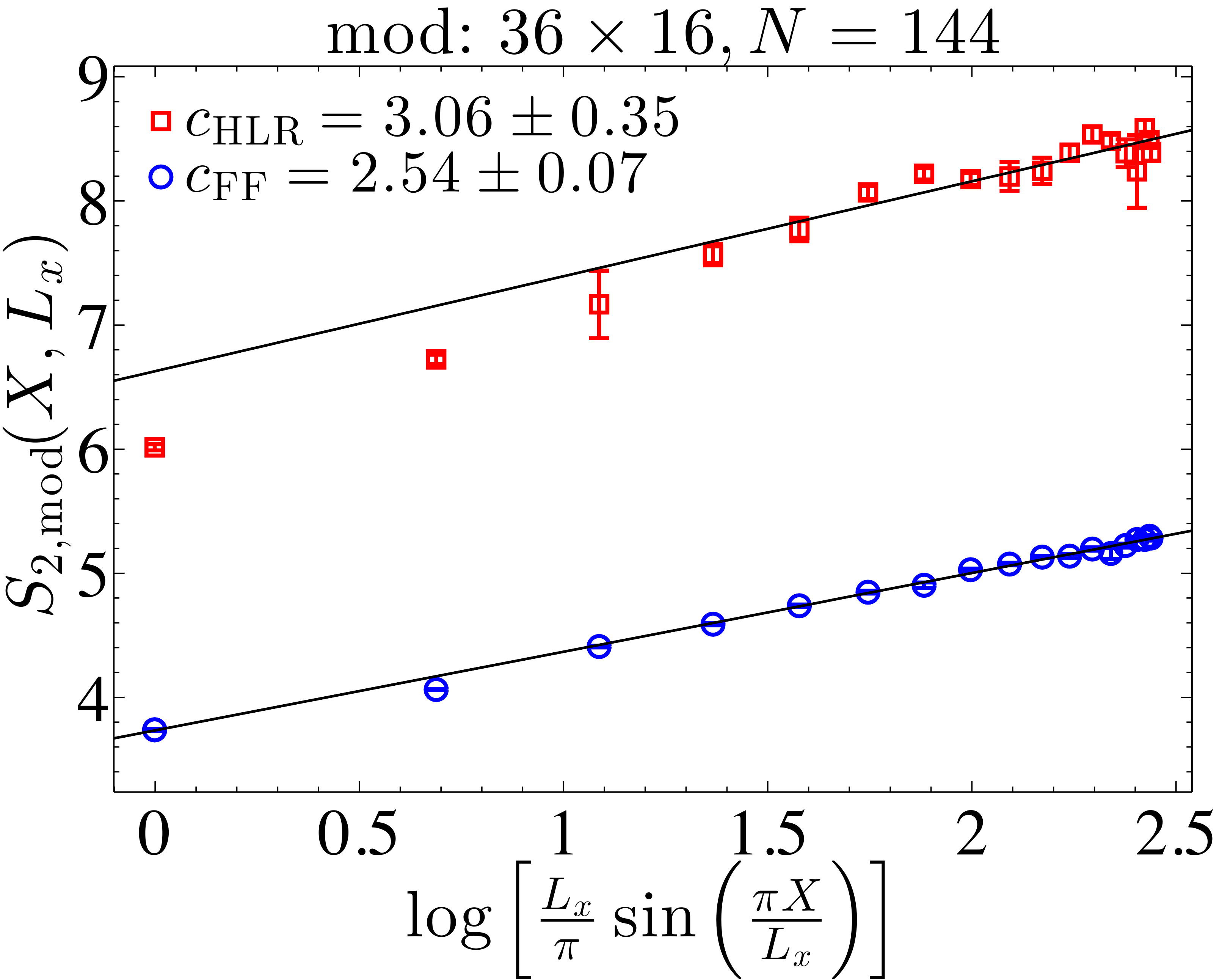}} \hfill
\subfigure{\includegraphics[width=0.28\textwidth]{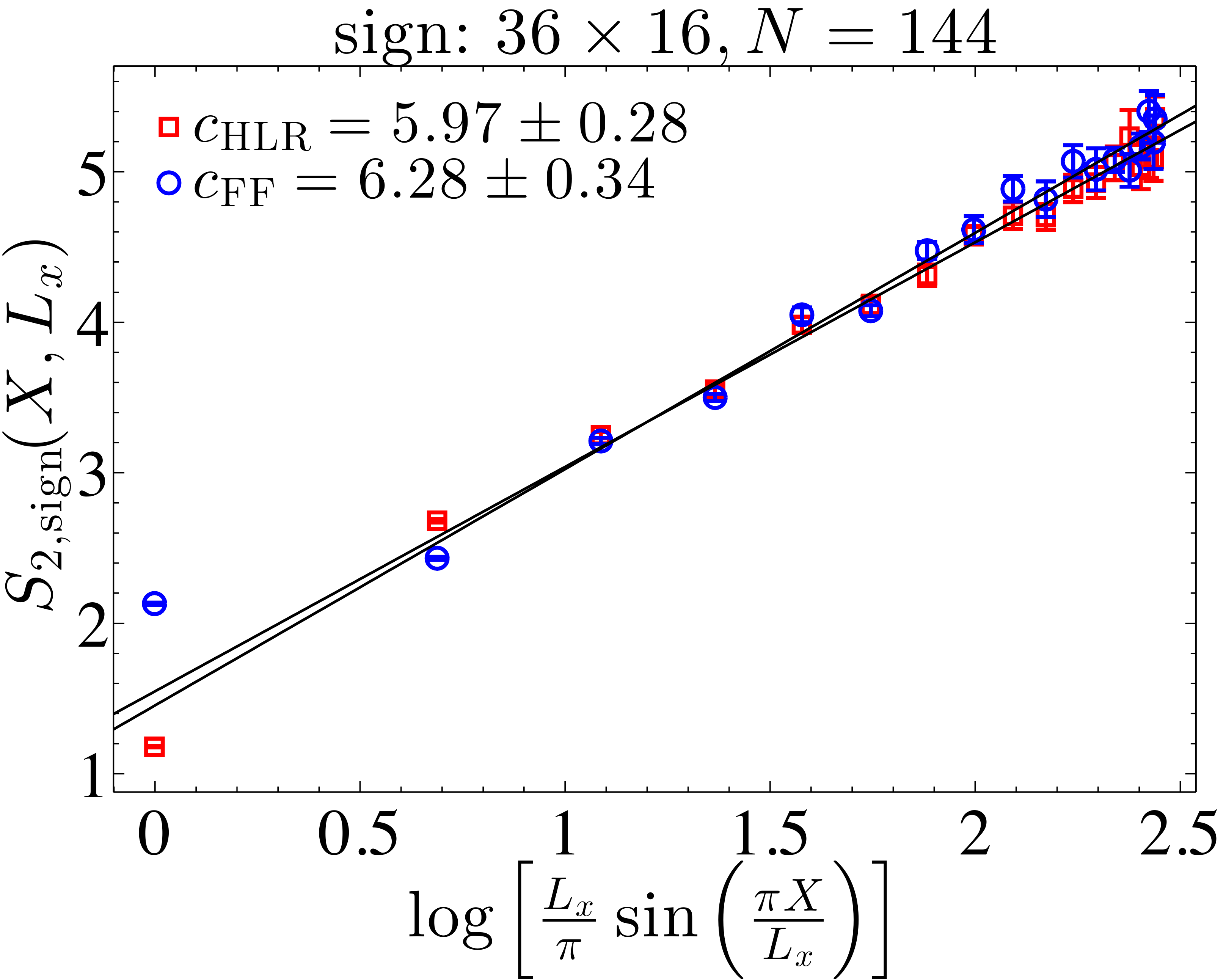}} \hfill
\hfill
}
\centerline{
\hfill
\subfigure{\includegraphics[width=0.28\textwidth]{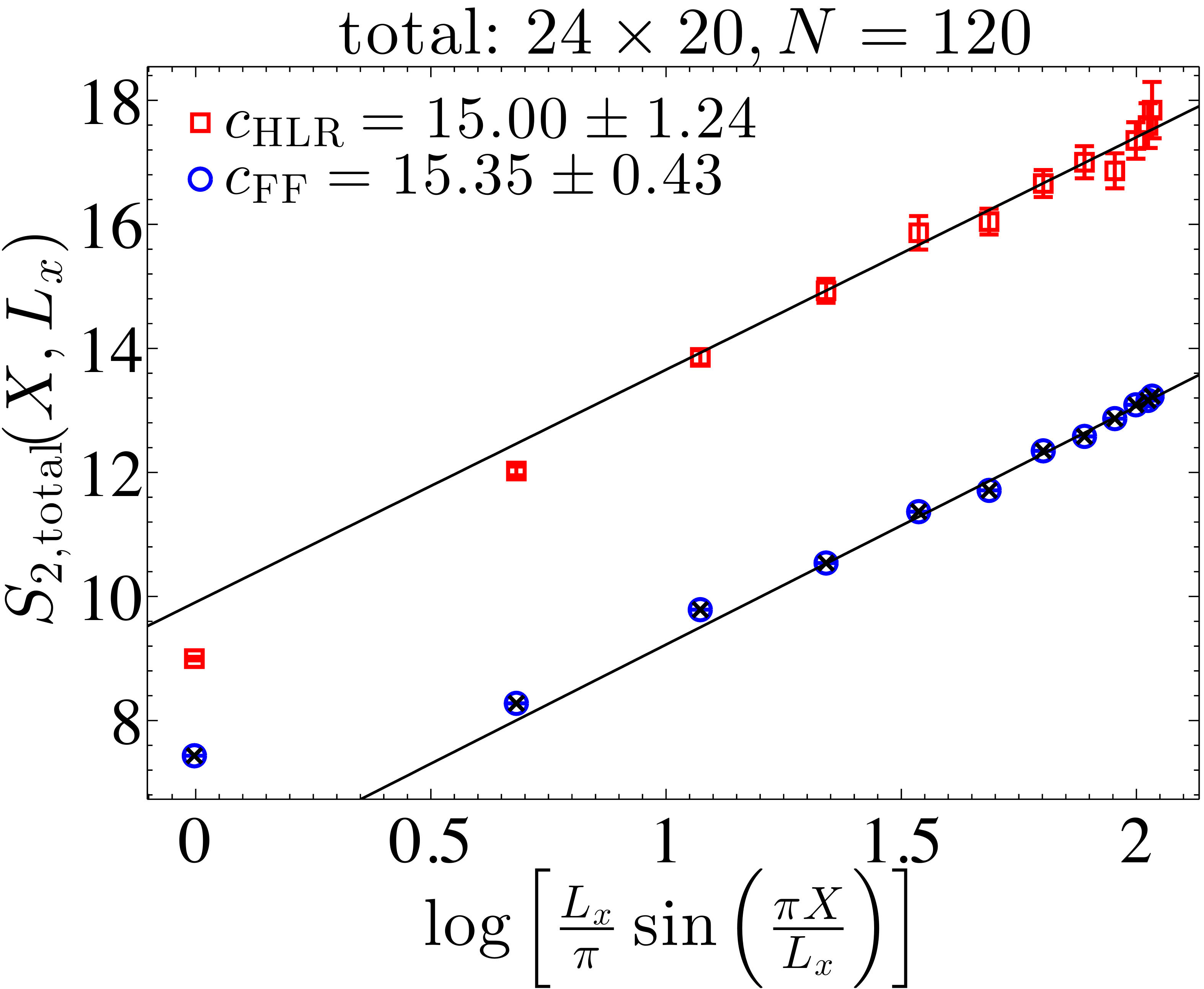}} \hfill
\subfigure{\includegraphics[width=0.28\textwidth]{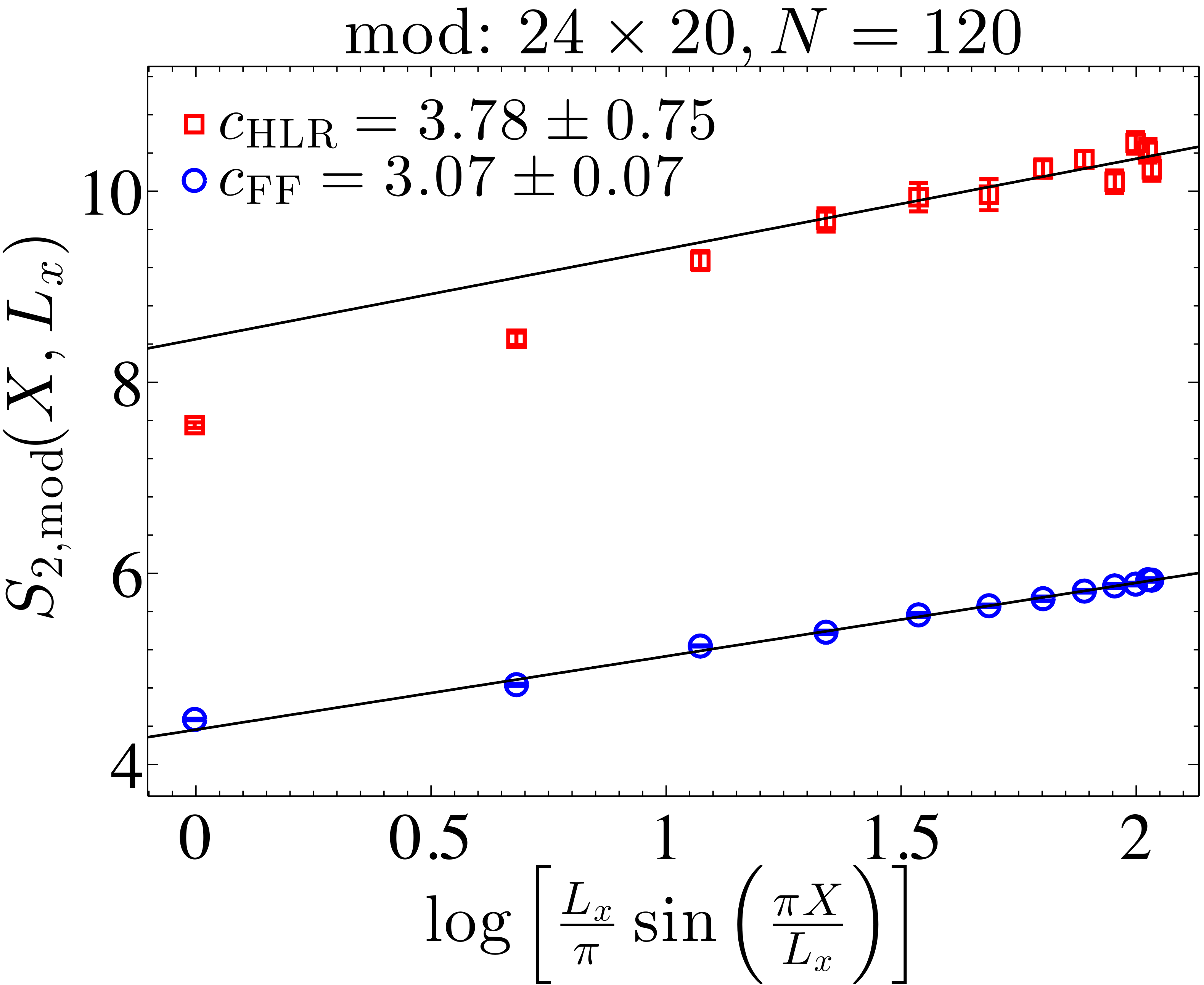}} \hfill
\subfigure{\includegraphics[width=0.28\textwidth]{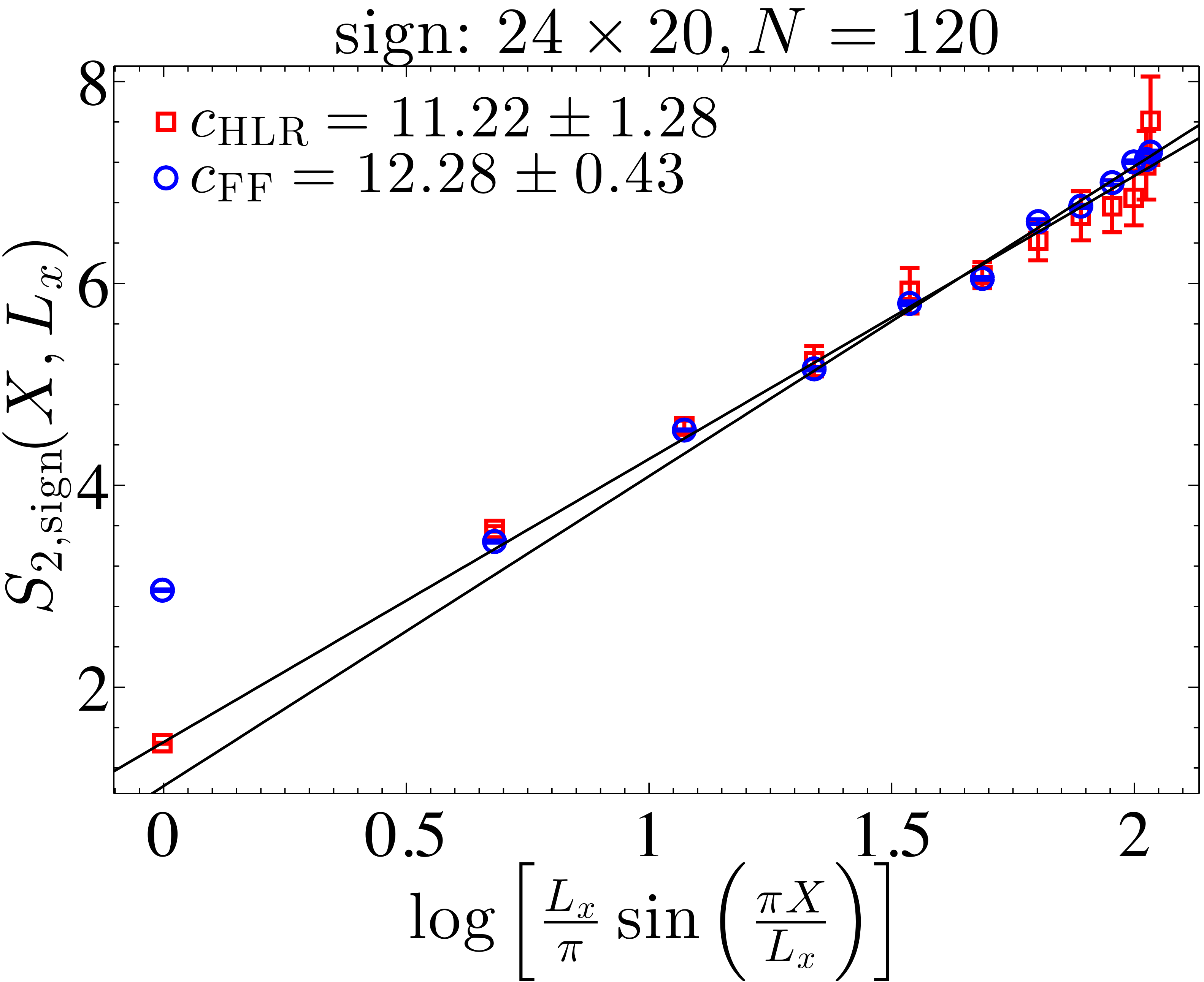}} \hfill
\hfill
}
\caption{Data analogous to Fig.~\ref{fig:strips} of the main text [$48\times12, N=144$ ($N_\mathrm{slices} = 7$)] but for the following systems from top row to bottom row:  $48\times4, N=48$ ($N_\mathrm{slices} = 3$); $48\times8, N=96$ ($N_\mathrm{slices} = 5$); $36\times16, N=144$ ($N_\mathrm{slices} = 9$); and $24\times20, N=120$ ($N_\mathrm{slices}$ is not well-defined for $L_x=24$, while $N_\mathrm{slices}=13$ for $L_y=20$ and large $L_x$).  The top three rows, as well as Fig.~\ref{fig:strips}, contain the data whose resulting fit parameters are plotted in Fig.~\ref{fig:fits}.}
\label{fig:all_strips}
\end{figure*}

While the two schemes are closely related, they require more or less independent implementations.  We have tested both implementations in the free fermion case, as well as against each other in the HLR case.  However, we found that the first scheme, i.e., the particle number trick, suffers from ergodicity problems when applied to the HLR states in quasi-1D geometries such as the 4-leg ladder.  We now prefer the second scheme as $(i)$ it generally works well in all geometries, and $(ii)$ it naturally explores all $n_A$ sectors according to their importance in the wavefunction instead of having to manually allocate computing time to each sector individually [cf.~Eq.~\eqref{eq:sectored}].  Still, the first scheme may be preferable in some instances.



\subsection{Strip geometry:  Summary and complete data sets}

In Fig.~\ref{fig:fits}, we summarize our strip geometry simulations for $L_y=4,8,12$, and 16 with $L_x=48,48,48$, and 36, respectively, all at $\rho=1/4$.  Figure~\ref{fig:fits}(a) shows the obtained central charge fit parameters $c_\mathrm{total}$, $c_\mathrm{mod}$, and $c_\mathrm{sign}$ for both the HLR and free fermion wavefunctions, while Fig.~\ref{fig:fits}(b) shows the corresponding intercepts $A$ [see Eq.~\eqref{eq:Cardy} of the main text].  The full data sets used to obtain these fits are shown in Fig.~\ref{fig:all_strips} (and in Fig.~\ref{fig:strips} of the main text).  In Fig.~\ref{fig:all_strips}, we also include data for a $24\times20, N=120$ system (bottom row).  This system has nearly unit aspect ratio and is far from the quasi-1D limit.  However, we still find the HLR and free fermion states to \emph{scale} nearly equivalently with $c_\mathrm{total}\approx15$, $c_\mathrm{mod}\approx3$, and $c_\mathrm{sign}\approx12$ in both cases.  ($N_\mathrm{slices}$ is not particularly well-defined here for $L_x=24$; see caption of Fig.~\ref{fig:all_strips}.)

For the $48\times4$, $48\times8$, and $48\times12$ systems, we excluded the smallest four $X$ values from the fits, while for $36\times16$ and $24\times20$, we excluded the smallest three.  Error bars in Fig.~\ref{fig:fits} and in the quoted $c$ values in Figs.~\ref{fig:strips} and \ref{fig:all_strips} are due to uncertainties in the fits only.

\end{document}